\def\R{{\rm I}\!{\rm R}}

\def\N{{\rm I}\!{\rm N}}

\def\carre{\hbox{\vrule \vbox to 7pt{\hrule width 6pt \vfill \hrule}\vrule }}
\parindent = 0cm

\font\un = cmbx10 at 14pt

\centerline {\un The classical limit of the time dependent}
\medskip
\centerline {\un  Hartree-Fock equation.}

\medskip
\centerline {\un II. The Wick symbol of the solution.}

\bigskip

 \centerline {\bf L. Amour, M. Khodja and J. Nourrigat}

 \bigskip

 \centerline {Laboratoire de Math\'ematiques,  EA4535}

\medskip

 \centerline {Universit\'e de Reims, U.F.R. Sciences Exactes et Naturelles}

\medskip

 \centerline {Moulin de la Housse - BP 1039, 51687 REIMS Cedex 2, France}

\medskip

 \centerline {and FR.CNRS.3399}

\vskip 1cm

The first aim of this paper is to derive under minimal hypotheses
Ehrenfest's theorem  concerning the time evolution of the Wick
symbol for a quantum observable (i.e, its average taken on coherent
states). This theorem (see [CTDL]) says that  the Wick symbol
follows the classical mechanics equations as the semiclassical
parameter $h$ goes to $0$. This result shall be precisely stated in
section 1 and proved in section 6 for observables that are not
pseudo-differential operators and consequently in the case where
this theorem cannot be deduced from Egorov's theorem. One may see
[R2], [CF], [CR1], [CR2], [S], [U1], [U2] for references concerning
coherent states and some of their applications and Ammari-Nier
[AN1][AN2] in the case of the infinite dimension.

\bigskip

Following our preceding work [AKN],  the second purpose here is to
study the time evolution of  the Wick symbol of a trace class
operator solution to the time dependent Hartree Fock equation
(TDHF). The work in [AKN] is devoted to the Weyl symbol of such of
an operator $\rho_h(t)$ under the hypothesis that the initial
operator $\rho_h(0)$ is a pseudo-differential trace class operator
belonging to the the  class of operators  studied by C. Rondeaux [R]
(actually, the class of operators in [R] is independent on $h$ and
the parameter $h$ is inserted here in order to consider
semi-classical operators). Under weaker hypotheses than those
assumed in [AKN] we shall study here the Wick symbol of such a
solution, instead of its Weyl symbol. A solution to (TDHF) is a
trace class operator and its Wick symbol is a function belonging to
$L^1(\R^{2n})$. We shall prove  that the time evolution of the Wick
symbol for a solution to the (TDHF) equation "tends to the time
evolution associated to the Vlasov equation" as the limit $h$ goes
to $0$. This limit shall be understood in the $L^1(\R^{2n})$ sense.
The precise statement of this result is written in section 2 (see
theorem 2.1) and its proof is given in section 7.

\bigskip

One of the tools used in order to established  both of these two
results is the approximation of a bounded operator in $L^2(\R^n)$,
(resp. of a trace class operator), verifying  rather weak
hypotheses, by pseudo-differential operators lying in the class of
Calderon-Vaillancourt [CV] (resp. in the class of C. Rondeaux [R]).
This process is similar to a convolution where translations are now
replaced by an action in the Heisenberg group. The details are found
in section 5.

\bigskip

In order to motivate the introduction of the Wick symbol it is given
in section 3 an explicit example of a trace class operator having a
Weyl symbol which is not in $L^1(\R^{2n})$. On the contrary, the
Wick symbol of a trace class operator is always in $L^1(\R^{2n})$
(this point is proved by C. Rondeaux [R]) together with all of its
derivatives (this fact is derived in section 4). Moreover,  an
asymptotic expansion of the Wick symbol for the product of two
operators when one of them is a semi-classical pseudo-differential
operator and when the other one is any trace class operator is
written in section 8. The bound for the remaining term of this
asymptotic expansion is effectuated in  the $L^1(\R^{2n})$ norm.
Note that an analysis for the commutator of an operator defined by
the Weyl calculus with another operator defined with the anti-Wick
formalism is found in R. Schubert [S] (theorem 4.1.12). The results
in section 8 below can be formally seen as the dual in some sense of
those in  R. Schubert. One may also see lemma 2.4.6 of N.Lerner
[L1]. In section 9 it is  obtained  an equation being
approximatively satisfied by the Wick symbol of a solution to (TDHF)
with an arbitrary small error term (see [DLERS] for the idea of such
an equation). Finally in section 10, it is observed  that the
convergence's result established in corollary 2.2 is not uniform on
$\R$ even if the two potentials are entirely vanishing.

\bigskip

{\bf 1. Statement of the result : the case of bounded operators and
Ehrenfest's theorem.}

\bigskip

In this section the time evolution of the system is associated to
the following Hamiltonian (which depend on $h>0$)

$$ \widehat H_h = - h^2 \Delta + V \leqno (1.1)$$
where $V$ is a $C^{\infty}$ real-valued function on $\R^n$ which is
bounded together with all of its derivatives. We also denote by
$\widehat H_h$ the unique self-adjoint extension of this operator.
Let $(A_h)_{h>0}$ be a family of bounded self-adjoint operators in
$L^2(\R^n)$. The operator $A_h(t)$ corresponding to the evolution of
the operator  $A$ at the time $t$ is
$$ A_h(t) = e^{i {t\over h} \widehat H_h} A_h e^{-i {t\over h} \widehat H_h}.  \leqno (1.2)$$
According to the standard statements of Ehrenfest's theorem (see
[CTDL]) the average of  $A_h(t)$ taken on coherent states is
supposed to approximatively follow  the classical mechanics's laws
as  $h$ tends to $0$ .

\bigskip

The coherent state centered at the point $X = (x , \xi)$ in
$\R^{2n}$ is the following function, depending on $h>0$
$$  \Psi_{X, h } (u) =  (\pi h)^{ -n/4}e^{-{| u-x|^2 \over
2h}} e^{{i \over h} u .\xi - {i \over 2h} x. \xi} \hskip 1cm X = (x
, \xi)\in \R^{2n}.  \leqno (1.3)$$
The average taken on $\Psi_{X h}$ of    a bounded operator $A$ in
${\cal H}= L^2 (\R^n)$ is
$$ \sigma_h^{wick}(A)(X) = < A \Psi_{X h} , \Psi_{Xh} >.  \leqno (1.4)$$
The function  $ \sigma_h^{wick}(A)$ is called the Wick symbol of
$A$. When $A$ is a bounded operator we shall notice  (see
proposition 4.1) that the function $ \sigma_h^{wick}(A)$ is
$C^{\infty}$ on $\R^{2n}$ and we shall give precise estimations on
its derivatives in terms of the parameter
 $h$ and of the norm of  $A$. Let us now specify in which sense does this function
 follows the classical mechanics equations under the limit $h$ tends to $0$.

\bigskip

Let $\varphi _t(x , \xi)= (q_t(x , \xi), p_t(x , \xi))$ denotes the
Hamiltonian flow of the function $H(x , \xi) = |\xi|^2 + V(x)$
starting at $(x , \xi)$, i.e., the solution to,
$$ q'_t (x , \xi) = 2 q(t , x , \xi) \hskip 2cm p'_t (x , \xi) =
- \nabla V ( q(t , x , \xi)) \leqno (1.5)$$
such that $q_0 (x , \xi) = x$, $p_0(x , \xi)=\xi$. One can say that
a function $f(x , \xi , t)$ "follows the classical mechanics laws"
when $f(X , t )= f( \varphi_t(X) , 0)$ for all $X\in \R^{2n}$ and
all $t\in \R$.

\bigskip

If the initial data
 $A_h$ is not assumed to be a semiclassical pseudo-differential operator (see D. Robert [R1]),
then the function $\sigma_h^{wick}(A_h(t))$ may not have a limit as
$h\rightarrow 0$. Nevertheless, we shall precisely state in which
sense this function "follows  the classical mechanics's equations
under the limit  $h$ tends to $0$". Our goal is to prove under
rather weak hypotheses and for all time $t \in \R$ that,
$$ \lim _{h\rightarrow 0}\Big | \sigma_h^{wick} (A_h(t) ) (X)   -
\sigma_h^{wick} (A_h(0) )  (\varphi _t(X)) \Big | = 0 \leqno (1.6)$$
In particular, this is not $\sigma_h^{wick} (A_h(t) ) (X)$ that has
a limit when $h\rightarrow 0$ but the difference between this
function and another one which follows the classical mechanics laws
and takes the same value as $\sigma_h^{wick} (A_h(t) ) (X)$ at time
$0$.

\bigskip

Let us now precisely state the hypotheses implying the validity of
the limit (1.6). We shall denote by $W^{m p} (\R^{2n})$ the space of
functions in $L^p (\R^{2n})$ having all of its derivatives up to
order $m$ in $L^p (\R^{2n})$  ($1 \leq p \leq +\infty $, $0 \leq m
\leq +\infty$).

\bigskip

First, let us recall how the Egorov theorem  is classically applied
the case of operators $A_h$ which are semiclassical
pseudodifferential operators. In this case,  $A_h$ is associated,
through the Weyl calculus, to a function $F_h$ in $W^{\infty \infty
}(\R^{2n})$, which  is  bounded in $W^{\infty \infty }(\R^{2n})$
independently of $h\in (0, 1]$. Namely, for all $f\in {\cal H}$,
$$ (A_h f) (x)  = ( 2 \pi h)^{-n} \int
_{\R^{2n}} F_h({x+y \over 2} , \xi )e^{{i\over h} (x-y).\xi} f(y)
 dy d\xi \leqno (1.7)$$
According to  Calderon-Vaillancourt [CV],  $A_h$ is bounded in
${\cal H}$. We agree to express (1.7) under the two following
expressions
$$ A_h = Op_h^{weyl} (F_h) \hskip 3cm F_h = \sigma_h^{weyl}(A_h)\leqno (1.8)$$
(The Weyl symbol of an  arbitrary bounded operator being a priori in
${\cal H}$ is a tempered distribution on $\R^{2n}$.) If $F_h$ is
either independent  of $h$ or admitting an asymptotic expansion in
the powers of $h$ as $h$ tends to $0$, Egorov's theorem shows that:
$$ \lim _{h \rightarrow 0} \Big ( \sigma_h^{weyl} (A_h(t)) (X) -
\sigma_h^{weyl} (A_h(0)) (\varphi_t(X))\Big )  = 0 \leqno (1.9)$$
In that case, the limit (1.6)  then follows. Indeed,
$$ \sigma_h^{wick} ( A_h (t)) = e^{{h\over 4}\Delta}
\sigma_h^{weyl} (A_h(t)) \leqno (1.10)$$
(Even if this result is standard, we shall nevertheless give a proof
in proposition 4.4.)  Besides, Egorov's theorem  also shows that, if
$A_h$ is written as in (1.7) with $F_h$ in $W^{\infty \infty
}(\R^{2n})$ and uniformly bounded  in  $h\in (0, 1]$ then
$\sigma_h^{weyl} (A(t)) $ is also a function in $W^{\infty \infty
}(\R^{2n})$, bounded uniformly on $h\in (0, 1]$.  As a consequence,
$$\lim _{h \rightarrow 0} \Big ( \sigma_h^{wick} (A_h(t)) (X) -
  \sigma_h^{weyl} (A_h(t)) (X) \Big ) = 0 \leqno (1.11)$$
for all $t\in \R$. In this case, the limit (1.6) then follows from
(1.9) and (1.11).

\bigskip

The assumption $A = Op_h ^{weyl}(F)$ with $F$ in $W^{\infty \infty
}(\R^{2n})$ is rather strong and may be expressed  in terms of
commutators in a standard way. Let $P_j (h)$ and $Q_j(h)$ be the
momemtum and the position operators,
$$ P_j (h) = {h\over i } {\partial \over \partial x_j} \hskip 2cm
Q_j (h) = x_j \leqno (1.12)$$
According to Beals characterization result [B], the fact that $A_h$
may be written as
 $A_h = Op_h ^{weyl}(F)$ with $F$ in  $W^{\infty \infty
}(\R^{2n})$ is equivalent to the fact that the commutators\ $(ad\
P(h) )^{\alpha} (ad \ Q(h))^{\beta } A$ are bounded (for all
multi-indexes $\alpha , \beta)$. Saying that $F$ (which may depend
on $h$) stays bounded in $W^{\infty \infty }(\R^{2n})$ is equivalent
to the fact,
$$  h^{-(|\alpha|+
|\beta|)} \Vert (ad\  P(h) )^{\alpha} (ad \ Q(h))^{\beta } A \Vert
_{{\cal L}({\cal H})} \leq M _{\alpha \beta}$$
with $M_{\alpha \beta}$ independent on $h$.

\bigskip

We are first concerned with the proof of the limit (1.6), which may
possibly be viewed as a form of Ehrenfest's theorem,  but under much
weaker hypotheses than in the case where $A_h$ is a semiclassical
operator. Namely, only single commutators (instead of iterated
commutators) of the operator $A_h$ with the operators $P_j(h)$ and
$Q_j(h)$ defined in (1.12) are assumed to be bounded operators. Our
estimates shall involve the following expression,
$$ I_h^{\infty} ( A) = {1 \over h}  \sum _{j=1}^n \Vert [P_j(h) , A  ]
 \Vert _{{\cal L}({\cal H})} + \Vert [Q_j(h) , A ]
 \Vert _{{\cal L}({\cal H})} \leqno (1.13) $$

\bigskip

The theorem below provides the inequality on which rely the proof of
Ehrenfest's theorem with weakened hypotheses.

\bigskip

{\bf Theorem 1.1} {\it For all operators $A$ in ${\cal L} ({\cal
H})$ such that the commutators $[P_j (h) , A]$ and $[Q_j (h) , A]$
are bounded operators in ${\cal H}$ ($1 \leq j \leq n$), the
operator $A_h (t)$ defined in (1.2) and the Hamiltonian flow
$\varphi_t$ defined in (1.5) satisfy
$$ \left \Vert \sigma_h^{wick } (A_h(t)) - \Big ( \sigma_h^{wick } (A) \Big ) \circ
\varphi_t    \right \Vert _{L^{\infty }(\R^{2n})}
 \leq C(t)   \sqrt{ h} I_h^{\infty} ( A) \leqno (1.14)$$
where $t \mapsto C(t)$ is a function defined on $\R$, bounded over
any compact set, depending only on $n$ and on  $V$ and where
$I_h^{\infty} ( A) $ is defined in (1.13).

}

\bigskip

In order that (1.6) can be  viewed as Ehrenfest's theorem it
suffices to replace $A$ by a family of operators $A_h$ such that the
right hand-side of   (1.13) tends to $0$ as $h$ goes to $0$. This
assumption is satisfied when $A_h = Op_h^{weyl} (F_h)$ with $F_h$ in
$W^{\infty \infty}(\R^{2n})$ and uniformly bounded in $h\in (0, 1]$.

 \vskip 2cm

{\bf 2. Statement of the result: the case of trace class operators
and the time dependent Hartree Fock equation.}

\bigskip

Before introducing the time dependent Hartree Fock equation (TDHF)
let us first specify the space to which the solution belongs. We
shall denote by ${\cal L}^1({\cal H})$ the space of all trace class
operators on ${\cal H} = L^2(\R^n)$. We shall denote by  ${\cal D}$
the subspace of operators $A$ in ${\cal L}^1({\cal H})$ such that
the commutator $[\Delta , A] $  is also in ${\cal L}^1({\cal H})$
where $\Delta$ is the Laplacian.

\bigskip

We consider two real-valued functions $V$ and $W$ in  $W^{\infty
\infty }(\R^{n})$. For all $h>0$ we shall say that a function $t
\rightarrow \rho_h (t)$ being $C^1$ from $\R$ into ${\cal L}^1({\cal
H})$ is a classical solution to the time dependent Hartree Fock
(according to the terminology of Bove da Prato Fano [BdF1,BdF2])
when this mapping is also continuous on $\R$ into ${\cal D}$ and if,
$$ ih {\partial \over \partial t} \rho_h (t) = - {h^2 }
[ \Delta , \rho_h(t ) ] + [ V_{q} (\rho_h(t)) , \rho_h (t )] \leqno
(2.1) $$
where $V_{q} (\rho_h(t))$ denotes the multiplication operator by the
function:
$$ V_{q} (x , \rho_h(t) ) =  V(x) + Tr ( W_x \rho_h (t )) \leqno (2.2)$$
where $W_x$ is the multiplication operator by the function $y
\rightarrow W(x -y)$.

\bigskip

We consider a family  $(\rho_h (t))_{h>0}$ of classical solutions to
(TDHF).  We suppose that the operator $\rho_h(0)$ is trace class,
self-adjoint $\geq 0$ with a trace equal to $1$. We set:
$$ u_h (X, t) = (2 \pi h)^{-n}  \sigma_h^{wick} ( \rho_h (t)) (X) \leqno (2.3)$$
As we shall see in section 4 this function lies in $W^{\infty
1}(\R^{2n})$, it is  $\geq 0$ and
$$ \int_{\R^{2n}} u_h (X , t) dX = 1 \leqno (2.4) $$
We denote by $v_h (X , t)$ the solution to the Vlasov equation,
$$ {\partial v_h \over \partial t }  + 2 \sum _{j=1}^n \xi_j {\partial
v_h \over \partial x_j}  -  \sum _{j=1}^n  {\partial V_{cl }(x ,
v_h(., t) ) \over \partial x_j} {\partial v_h \over \partial \xi _j}
=0 \leqno (2.5)$$
such that
$$ v_h(X , 0) = u_h (X , 0) \leqno (2.6) $$
We have set,
$$V_{cl }(x , v_h(., t) ) =  V(x) +  \int _{\R^{2n}} W(x -y)
 v_h(y , \eta , t) dy d\eta \leqno (2.7) $$
The function $v_h (., t)$ is itself in $L^1(\R^{2n})$, it is $\geq
0$ and its integral equals to $1$.

\bigskip

The counterpart of Ehrenfest's theorem for a family of solutions to
(TDHF) consists into saying that the function
 $u_h(., t)$ defined in
(2.3) satisfies Vlasov equation "under the limit $h$ tends to $0$".
This point of view may specified in different ways. In this section
our aim is to compare as in theorem 1.1 the functions $u_h(., t)$
and $v_h(., t)$ and to show that, under suitable assumptions on
$\rho_h(0)$, their difference in norm tends to $0$. Since we are
concerned with trace class operators it makes sense that the norm
involved in the estimation of the difference  $u_h(., t) -v_h(., t)$
is the  $L^1(\R^{2n})$ norm. A first answer to this issue is given
in our preceding article (preprint) [AKN] where we assume that the
operator
 $\rho_h(0)$ is a
pseudo-differential operator belonging to a class of operators
introduced by C. Rondeaux [R] (the only modification here being the
insertion of the parameter $h$). The article [AKN] also gives an
asymptotic expansion at any order of $u_h(., t) -v_h(., t)$. We
consider here this problem with a weaker hypothesis  than the one in
[AKN]. We are only assuming that all the commutators $[P_j(h) ,
\rho_h(0)  ]$ and $[Q_j(h) , \rho_h(0)  ]$ are trace class
operators. All these estimates shall use the expression,
$$ I_h^{tr} ( \rho_h(0)) = {1 \over h}  \sum _{j=1}^n \Vert [P_j(h) , \rho_h(0)  ]
 \Vert _{{\cal L}^1({\cal H})} + \Vert [Q_j(h) , \rho_h(0)  ]
 \Vert _{{\cal L}^1({\cal H})} \leqno (2.8) $$

\bigskip

{\bf Theorem 2.1.} {\it Let  $(\rho_h (t))_{h>0}$ be a family of
classical solutions to (TDHF) with real-valued potentials $V$ and
$W$ in $W^{\infty \infty }(\R^{2n})$. We suppose that the operator
$\rho_h(0)$ is trace class, self-adjoint $\geq 0$, with a trace
equal to $1$. We assume that all the commutators $[P_j(h) ,
\rho_h(0)  ]$ and $[Q_j(h) , \rho_h(0) ]$ are trace class operators.
Then, there exists a function  $t \mapsto C(t)$, bounded on any
compact set of $\R$ such that,
for all $h\in (0, 1]$ and for all $t\in \R$,
$$\Vert  u_h(., t) - v_h(., t) \Vert _{L^1(\R^{2n})}
\leq C(t)\sqrt {h}   I_h^{tr}( \rho_h(0))e^{C(t) I_h^{tr}
(\rho_h(0))} \leqno (2.9) $$

}

\bigskip

{\bf Corollary 2.2.}  {\it Under the assumptions of theorem 2.1, if
$I_h^{tr}(\rho_h(0))$ remains bounded when $h$ tends to $0$, then
we have,
$$ \lim _{h \rightarrow 0}\Vert  u_h(., t) - v_h(., t) \Vert
_{L^1(\R^{2n})} = 0 \leqno (2.10)$$ for all $t\in \R$.
}

\bigskip

If $\rho_h(0)$ is a pseudo-differential operator written as
$\rho_h(0) = (2\pi h)^n Op_h^{weyl} (F_h)$ with $F_h$ in $W^{\infty
1}(\R^{2n})$ and uniformly bounded in $h$ then   the assumption in
corollary 2.2 is always satisfied according to the analogous result
of the Beals's characterization  given in [R]  (see [AKN] for
dependence on the parameter $h$).

\bigskip

Theorem 2.1 shall be proved in section 7. An other way to give an
insight into Ehrenfest's theorem consists into proving that the
function $u_h(X , t)$ satisfies an equation that formally shrink to
the Vlasov equation when  $h$ tends to $0$. It is this point of view
that we shall develop in section 9.

\bigskip

{\bf 3. A counter-example.}

\bigskip

By making explicit an idea of  C. Rondeaux [R] we shall give an
example of a trace class operator having a Weyl symbol not being in
$L^1 (\R^{2n})$. Let us mention before some properties for the Weyl
symbol of a trace class operator that lead us to the choice of the
counter-example.

\bigskip

{\bf Proposition 3.1.} {\it If $A$ is trace class,  then its Weyl
symbol is a continuous function on $\R^{2n}$ going to $0$ at
infinity and also belonging in $L^2 (\R^{2n})$. If this function is
also $\geq 0$ then it is necessarily in $L^1(\R^{2n})$.

}

\bigskip

{\it Proof.} If $A$ is trace class then it is also Hilbert-Schmidt
and it is well-known that its Weyl symbol belongs to $L^2(\R^{2n})$
(cf D. Robert [R1]). This symbol is bounded since when $A$ is trace
class we have,
$$ \sigma_h^{weyl} (A) (X) = 2^n Tr ( A \circ \Sigma_{Xh}) \hskip 2cm X \in \R^{2n}
\leqno (3.1) $$
The fact that it is continuous and that it is going to $0$ at
infinity is easily verified using (3.1) for an operator written as
$ u \rightarrow A(u)= < u , \varphi > \psi$, where $\varphi$ and
$\psi$ are in ${\cal S}(\R^n)$. One then concludes this point by
density, first for any finite rank  $A$, then for any trace class
operator since the set of finite rank operators is dense in  ${\cal
L}^1({\cal H} )$. For all functions  $F$ in $W^{\infty 1}(\R^{2n})$
and  $G $ in $W^{\infty \infty }(\R^{2n})$  it is well-known (cf D.
Robert [R1]) that,
$$ Tr (Op_h^{weyl}(F)  \circ Op_h^{weyl}(G) ) = (2\pi h)^{-n} \int _{\R^{2n}}
F(X) \ G(X) \ dX  \leqno (3.2)$$
Let $A$ be a trace class operator and $G$ be a $C^{\infty}$ function
on  $\R^{2n}$ with compact support. From theorem 5.2 there exists a
sequence of functions $(F_j)$ in $W^{\infty 1}(\R^{2n})$ such that
the sequence of operators $Op_h^{weyl}(F_j)$ converges to $A$ in
${\cal L}^1({\cal H})$, implying from (3.1) that $F_j$ converges
uniformly to $\sigma_h^{weyl}(A)$. It is then deduced that,
$$ Tr (A  \circ Op_h^{weyl}(G) ) = (2\pi h)^{-n} \int _{\R^{2n}}
\sigma_h^{weyl}(A) (X)  \ G(X) \ dX  \leqno (3.3)$$
We replace $G$ by an increasing sequence  $G_N$ of $C^{\infty}$
functions
 on $\R^{2n}$ with compact support and converging pointwise
to 1 when $N$ goes to $+\infty$, the functions $|\partial
_x^{\alpha}
\partial _{\xi}^{\beta} G_N|$  being all uniformly bounded
 on $N$. When $\sigma_h ^{weyl}(A) \geq 0$ we deduce from (3.3) that,
$$ 0 \leq (2\pi h)^{-n} \int _{\R^{2n}}
 \sigma_h ^{weyl}(A) (X) \ G_N(X) \ dX \leq \Vert A \Vert_{{\cal
 L}^1 ({\cal H})} \ \Vert Op_h^{weyl}(G_N) \Vert_{{\cal L} ({\cal
 H})}$$
According to Calderon-Vaillancourt the right hand-side remains
bounded as $N$ tends  $+\infty$. If the function $\sigma_h
^{weyl}(A) $ is $\geq 0$ then it is therefore  in $L^1(\R^{2n})$.

\hfill \carre

\bigskip

Proposition 3.1 and the results of C. Rondeaux [R] suggest the
construction of an example of a trace class operator having a Weyl
symbol not being in $L^1 (\R^{2n})$. We may assume that  $h=1$. Let
$\alpha $ be a real number such that ${n \over 2 } < \alpha \leq n$.
We define the operator $P$ by:
$$ P = Op_1^{weyl} (p) \hskip 2cm p(x , \xi) = { e^{2i x .\xi } \over
(1+ |x|^2 + |\xi|^2)^{\alpha}} \hskip 2cm (x , \xi)\in \R^{2n}.
\leqno (3.4) $$
The function $p$ is not in $L^1 (\R^{2n})$. For all $\lambda
>0$ we define an operator $A_{\lambda}$ by
$$ A_{\lambda } = Op_1^{weyl} (a_{\lambda}) \hskip 2cm a_{\lambda }
(x , \xi) =e^{2i x \xi  - \lambda (|x|^2 + |\xi|^2)}.$$
We have:
$$ P = {1 \over \Gamma (\alpha ) }  \int _0^{\infty} e^{-\lambda } \lambda ^{\alpha - 1} A_{\lambda }
d \lambda  \leqno (3.5) $$
provided that we verify the convergence. From definition (1.7) the
integral kernel of $A_{\lambda}$ is the function $K_{\lambda}$
defined by:
$$ K_{\lambda}(x , y) = (2 \pi)^{-n} \int _{\R^n} e^{i (x-y).\xi}
a_{\lambda} \left ( {x+y\over 2} , \xi \right ) d\xi. $$
An explicit computation shows that:
$$ K_{\lambda } (x , y) =  (2\pi  \lambda ) ^{-n/2} \ e^{ - (
a(\lambda)  |x|^2 + b(\lambda) |y|^2 + 2 c(\lambda) x.y)} \hskip 2cm
a (\lambda) = {\lambda \over 4} + {1 \over \lambda} \hskip 2 cm
b(\lambda) = c (\lambda )= {\lambda \over 4}. $$
We can express the operator  $A_{\lambda }$ as a product,
$$A_{\lambda } = (2 \pi \lambda b(\lambda ) )^{-n/2}  \
T_{a(\lambda )} \circ B_{\lambda } \circ T_{b(\lambda )}^{-1} \circ
S \leqno (3.6) $$
where for all $a>0$, $(T_af) (x) = f( x \sqrt {a})$, $Sf (x) = f(
-x)$ and:
 $$ (B_{\lambda } f ) (x) = \int _{\R^n } e^{ - (|x|^2 + |y|^2) + 2 \mu
 (\lambda ) x .y } f(y) dy \hskip 2cm \mu (\lambda) = { c(\lambda)
 \over \sqrt { a(\lambda ) b(\lambda) }}.  \leqno (3.7)$$
The operator $B_{\lambda }$ is self-adjoint $\geq 0$ (since $0 < \mu
(\lambda) < 1$ and $B(\lambda)$ may be identified,  up to a
multiplicative constant, to  the exponential of an harmonic
oscillator). We then see:
$$ \Vert B_{\lambda } \Vert _{{\cal L}^1 ({\cal H})} = {\rm Tr}
(B_{\lambda}) = \int _{\R^n} e^{ - 2(1 -\mu
 (\lambda ))  |x|^2  } dx =  \left ( {
\pi \over 2 ( 1 - \mu (\lambda ))}\right )^{n/2}.   $$
From (3.6) we have,
$$ \Vert A_{\lambda } \Vert _{{\cal L}^1 ({\cal H})} \leq
(2 \pi  \lambda b(\lambda ) )^{-n/2} \  \Vert T_{a(\lambda )}\Vert
_{{\cal L} ({\cal H})} \  \Vert B_{\lambda } \Vert _{{\cal L}^1
({\cal H})}\ \Vert T_{b(\lambda )}^{-1} \Vert _{{\cal L} ({\cal
H})}.
$$
The right hand-side is polynomially increasing as  $\lambda
\rightarrow +\infty $ and it is ${\cal O} (\lambda ^{-n/2})$ when
$\lambda$ tends to $0$. With the choice of  $\alpha > {n\over 2}$,
the integral (3.5) converges in norm in ${\cal L}^1 ({\cal H})$ and
it properly defines a trace class operator $P$,  the Weyl symbol of
which is the function $p$ defined in (3.4), and  is not belonging to
$L^1(\R^{2n})$.

\bigskip

For  trace class operators,  it is natural  to give a variant of
Ehrenfest's theorem where the limit is understood in the
$L^1(\R^{2n})$ sense. The above counter-example shows that the Weyl
symbol is not suitable for this purpose without any supplementary
hypothesis. However, proposition 4.2 below shows that the Wick
symbol of a trace class operator is $C^{\infty }$ from $\R^{2n}$
into $L^1(\R^{2n})$  together with all of its derivatives.

\bigskip

{\bf 4. Differentiability of the Wick symbol.}

\bigskip

{\bf Proposition 4.1.} {\it Let $A$ be a bounded operator in ${\cal
H}$. Then the Wick symbol of  $A$, namely, the function
$\sigma_h^{wick} (A)$ defined on $\R^{2n}$ by (1.4), is a $C^{\infty
}$  function on $\R^{2n}$, bounded together with all of its
derivatives.  For each multi-index $\alpha $ there exists $C_{\alpha
}$ such that:
$$ \Vert  \nabla ^{\alpha}
\sigma_h^{wick} (A)  \Vert _{L^{\infty }(\R^{2n})}  \leq C_{\alpha }
h^{-|\alpha| /2} \ \Vert A \Vert _{{\cal L} ({\cal H}) }.  \leqno
(4.1)$$
}

\bigskip

{\it Proof.} We shall use the following function,
$$ S_h  (A) (X , Y) = { < A \Psi_{Xh}, \Psi_{Yh}> \over < \Psi_{Xh}, \Psi_{Yh}>
},  \leqno (4.2)$$
defined on $\R^{2n} \times \R^{2n}$,  where the functions $\Psi_{X
h}$ are given in (1.3). Direct computations shows that:
$$< \Psi_{Xh} , \Psi_{Yh}> =  e^{-{1\over 4h}|X-Y|^2 +
{i \over 2h} {\rm Im} (X .\overline Y )}.  \leqno (4.3) $$
In particular,
$$ | < \Psi_{Xh} , \Psi_{Yh}>| = e^{- {1 \over 4h} |X-Y|^2} \hskip  3cm \Vert \Psi_{Xh}\Vert =
1. \leqno (4.4)$$
Consequently:
$$
  \Big |
S_h (A) (X , Y ) \Big |  =  e^{{1\over 4h} |X-Y|^2} |< A \Psi_{Xh} ,
\Psi_{Y h} > |  \leq e^{{1\over 4h} |X-Y|^2} \ \Vert A \Vert _{{\cal
L} ({\cal H}) }.  \leqno (4.5)
$$
An important property verified by coherent states is that:
$$ < f , g> = (2 \pi h)^{-n} \int _{\R^{2n}} < f , \Psi _{X h} > \ <
\Psi _{X  h} , g > \ dX \leqno (4.6) $$
for all $f$ and $g$ in ${\cal H}$. Applying this equality, we see
that:
$$\sigma _h ^{wick} (A) (X) = S_h (A) (X , X ) = (2\pi h)^{-2n} \int _{\R^{4n}} {\cal B}_h ( X ,
U , V , X) S_h ( A ) (U , V ) dU dV \leqno (4.7) $$
where for all $(X , U , V , Y)$ we have set:
$${\cal B}_h ( X , U , V , Y ) = { < \Psi_{Xh} , \Psi _{U h} > < \Psi_{Uh} , \Psi _{V h}
> < \Psi_{V h} , \Psi _{Y h} > \over < \Psi_{Xh} , \Psi _{Y h} > }.  \leqno (4.8) $$
From (4.3) we have:
$${\cal B}_h ( X , U , V , X) = e^{-{1 \over 2h} ( X- V) ( \overline X - \overline U) -
{1 \over 2h} |U - V|^2}. \leqno (4.9) $$
We verify that:
$$ h^{m/2} | \nabla _X^{\alpha}{\cal B}_h ( X , U , V , X) |
\leq C_m K_m (X , U , V , h)e^{ -{1 \over 4h}  |U - V|^2 }$$
where
$$ K_m (X , U , V , h) = \left ( 1 + {|X - U| + |X-V|\over \sqrt h} \right )^m \ e^{ -{1
\over 4h} ( | X - U|^2 +  |V-X|^2 )}.
$$
Consequently:
$$  h^{m/2} \Big | \nabla _X^{m}\sigma _h ^{wick} (A) (X)
\Big | \leq  C_m (2\pi h)^{-2n} \int _{\R^{4n}}     |< A \Psi_{U h}
, \Psi_{V  h} > | K_m (X , U , V , h) \  \ dU dV.  \leqno (4.10)$$
By bounding from above  $ |< A \Psi_{U h} , \Psi_{V  h} > | $ by the
norm of $A$ we then obtain (4.1).

\bigskip

{\bf Proposition 4.2.}  {\it If $A$ is in ${\cal L}^1({\cal H})$,
then its   Wick symbol $\sigma_h (A)$ belongs to $W^{\infty 1}
(\R^{2n})$. For each multi-index $\alpha $ there exists $C_{\alpha
}>0$ such that,
$$(2\pi h)^{-n} \Vert \nabla ^{m}  \sigma_h
^{wick} (A)  \Vert _{L^1(\R^{2n})} \leq C_{m} h^{-{m\over 2}} \Vert
A \Vert _{{\cal L}^1({\cal H})}.  \leqno (4.11)$$

}

\bigskip

Inequality (4.11) is proved when $m = 0$ by C. Rondeaux [R].

\bigskip

{\it Proof.} Similarly to proposition 4.1, we have inequality
(4.10). Integrating with respect to $X$ we obtain:
$$h^{m/2} (2\pi h)^{-n} \Vert \nabla ^{m}  \sigma_h
^{wick} (A)  \Vert _{L^1(\R^{2n})} \leq C_{m}(2\pi h)^{-2n}
 \int _{\R^{4n}}     |< A \Psi_{U h}
, \Psi_{V  h} >  G_m \left ( {U-V \over \sqrt h} \right ) \  \ dU
dV$$
$$ G_m (U) = (1 + |U|)^m e^{- {|U|^2 \over 8}}. $$
Since the function $G_m $ is in $L^1(\R^{2n})$,  proposition 3.2 is
a consequence of the following lemma.

\bigskip

{\bf Lemma 4.3.} {\it Let $A$ be a trace class operator and $G$ be a
function in $L^1(\R^{2n})$. Then we have:
$$ (2\pi h)^{-2n}  \int _{\R^{4n}}    \Big  |< A \Psi_{X h}
, \Psi_{Y  h} >   G\left ( {X-Y \over \sqrt h} \right )  \Big |  \
dX dY \leq (2\pi)^{-n}  \Vert G \Vert _{L^1(\R^{2n})} \Vert A \Vert
_{{\cal L}^1({\cal H})}.  \leqno (4.12)$$
}

\bigskip

{\it Proof.} We can write $A = B_1 B_2$ where $B_1$ and $B_2$ are
two Hilbert-Schmidt operators. Using the fundamental indentity (4.6)
verified by the coherent states we see that, for all $X$ and $Y$ in
$\R^{2n}$,
$$ < A \Psi_{Xh} , \Psi_{Y h} > = < B_2 \Psi_{Xh} , B_1^{\star}  \Psi_{Y h} >
= (2 \pi h)^{-n} \int _{\R^ {2n}} u_{Zh} (X)  \  v_{Zh} (Y) \  dZ$$
where we have set $u_{Zh } (X) = < B_2 \Psi_{Xh} , \Psi_{Zh}> $ and
$v_{Zh } (X) = < \Psi_{Zh} ,  B_1^{\star} \Psi_{X h} > $. Let $I_h$
be the left hand-side of (4.12). From Schur's lemma:
$$ I_h \leq (2\pi h)^{-3n} h^n  \Vert G \Vert _{L^1(\R^{2n})} \  \int _{\R^{2n}} \Vert u_{Zh} \Vert
 _{L^2(\R^{2n})} \ \Vert v_{Zh} \Vert _{L^2(\R^{2n})} dZ. $$
From (4.6), we have $\Vert u_{Zh} \Vert
 _{L^2(\R^{2n})}  = (2 \pi h)^{n/2} \Vert B_2^{\star} \Psi_{Zh} \Vert  $ and $ \Vert v_{Zh} \Vert
 _{L^2(\R^{2n})}  = (2 \pi h)^{n/2}  \Vert B_1 \Psi_{Zh} \Vert$. Consequently,
$$ I_h \leq  (2\pi h)^{-2n} h^n  \Vert G \Vert _{L^1(\R^{2n})} \    \int _{\R^{2n}} \Vert
B_1 \Psi_{Zh} \Vert \ \Vert B_2^{\star} \Psi_{Zh} \Vert \ dZ. $$
From the basic indentity (4.6) for coherent states,
$$(2\pi h)^{-n} \int _{\R^{2n}}  \Vert
B_j \Psi_{Zh} \Vert ^2 dZ = (2\pi h)^{-n} \int _{\R^{2n}}  < B_j
^{\star} B_j  \Psi_{Z h} , \Psi_{Zh} > \ dZ$$
$$ = {\rm Tr } ( B_j ^{\star } B_j) = \Vert B_j\Vert _{{\cal L}^2 ({\cal H})}^2$$
where $ \Vert B_j\Vert _{{\cal L}^2 ({\cal H})}$ is the
Hilbert-Schmidt norm of  $B_j$. Therefore,
$$ I_h  \leq   (2\pi)^{-n}   \Vert G \Vert _{L^1(\R^{2n})} \ \Vert B_1 \Vert _{{\cal
L}^2 ({\cal H})}\ \Vert B_2 \Vert _{{\cal L}^2 ({\cal H})}. $$
Taking the infimum over all $A$ written as $A = B_1 B_2$ we then
obtain (4.12).

 \hfill \carre

\bigskip

Let us finally recall the following proposition even if it is
standard.

\bigskip

{\bf Proposition 4.4.} {\it The Wick symbol $\sigma_h ^{wick}(A)$ of
an operator $A$ is related to its Weyl symbol $\sigma_h ^{weyl}(A)$
by
$$\sigma_h ^{wick}(A) = e^{{h\over 4}\Delta} \sigma_h ^{weyl}(A) \leqno (4.13)$$
where $\Delta$ is the Laplacian on $\R^{2n}$.

}
\bigskip

Indeed, setting $F_h = \sigma_h ^{weyl}(A)$, the expression in
definition (1.7) for the Weyl calculus may be written as:
$$ A =  (\pi h)^{-n}  \int_{\R^{2n}} F_h(Y) \Sigma _{Yh}
dY, \leqno (4.14)$$
where, for all $Y = (y , \eta)$ in $\R^{2n}$,   $\Sigma _{Y h}$ is
the operator ("symmetry operator") defined by,
$$ (\Sigma _{Y h} f) (u) = e^{{2i\over h} (u-y)\xi } f( 2y - u)
\hskip 1cm  Y = (y , \eta )\in \R^{2n}.  \leqno (4.15) $$
A direct computation shows that:
$$ \sigma_h^{wick} (  \Sigma_{Yh} ) (X) = < \Sigma_{Yh}  \Psi_{X h} , \Psi_{Xh}>
= e^{-{|X-Y|^2 \over h}}.  \leqno (4.16)$$
Equality (4.13) then follows.

\bigskip

{\bf 5. Pseudo-differential approximation.}

\bigskip

The class of operators having a Weyl symbol in $W^{\infty
1}(\R^{2n})$  (introduced by C. Rondeaux) is dense in the set of
trace class operators similarly to the fact that  $W^{\infty
1}(\R^n)$ is dense in $L^1(\R^{n})$.  The approximation process has
a strong connection with convolutions and may probably be also
applied to Schatten's classes. This process is often employed in
representation theory. However, as for a function in
$L^{\infty}(\R^{n})$ needs to be continuous in order to be in the
closure of $W^{\infty \infty }(\R^n)$, a bounded operator needs
additional hypotheses in order to be in the closure of the class of
Calderon-Vaillancourt for operators.

\bigskip

For all $X = ( x , \xi) $ in $\R^n$ and  for all $h>0$ let $W_{x ,
\xi , h}$ be the operator defined by
$$( W _{ X , h  } f) =  ( W _{ x , \xi , h  } f) (u) = f(u - x ) e^{ {i\over h}
u . \xi - {i\over 2h} x. \xi} \leqno (5.1)$$
for all $f\in L^2(\R^n)$. It is a common representation of the group
of Heisenberg. Thus, the coherent state $\Psi_{Xh}$ verifies,
$$ \Psi _{X , h} = W _{X , h } \Psi_{0 , h} \leqno (5.2) $$
We have, for all $X$ and $Y$ in $\R^{2n}$
$$ W _{ X , h  }  W _{ Y , h  } = e^{ {i\over 2h} \sigma (X , Y) } W _{ X+Y  , h  }
 \leqno (5.3)$$
where $\sigma $ is the symplectic form $\sigma (( x , \xi) , (y ,
\eta)) = y.\xi - x .\eta $. For all operators $A$ in ${\cal L}({\cal
H})$ and for all $h>0$ let us define,
$$ {\cal T}_h A = ( \pi h)^{-n} \int _{\R^{2n}} e^{ - {|X|^2 \over h} }
W _{X , h } A W_{X , h}^{\star } \ dX \leqno (5.4) $$

\bigskip

We begin with the case of bounded and trace class operators
satisfying some hypotheses involving their commutators with position
and momentum operators.

\bigskip

{\bf Theorem 5.1.} {\it a) We have:
$$ \Vert {\cal T}_h A \Vert_{{\cal L}({\cal H}) } \leq
 \Vert A \Vert_{{\cal L}({\cal H}) } \leqno (5.5)$$
for all operators $A$ in ${\cal L}({\cal H})$ and for all $h>0$.

\smallskip

b) When the commutators $[ P_j (h) ,  A]$ and  $[ Q_j (h), A]$ are
bounded operators we have:
$$ \Vert A -  {\cal T}_h A \Vert_{{\cal L}({\cal H}) } \leq
{ C \over \sqrt h}  \sum _{j= 1}^n \Vert [ P _j (h) ,   A]
\Vert_{{\cal L}({\cal H}) } + \Vert [Q _j (h),   A] \Vert_{{\cal
L}({\cal H}) }.  \leqno (5.6)$$
c) When the operator $A$ is trace class together with the
commutators $[ P_j (h) , A]$ and  $[ Q_j (h), A]$, we have:
$$ \Vert A -  {\cal T}_h A \Vert_{{\cal L}^1({\cal H}) } \leq
{ C \over \sqrt h}  \sum _{j= 1}^n \Vert [ P _j (h) ,   A]
\Vert_{{\cal L}^1({\cal H}) } + \Vert [Q _j (h),   A] \Vert_{{\cal
L}^1({\cal H}) }.  \leqno (5.7)$$
d) The Wick symbol of the operators $A $ and $ {\cal T}_h A$ are
related with:
$$ \sigma_h^{wick} ({\cal T}_h A ) = e^{{h\over 4 } \Delta }
\sigma_h^{wick} ( A ).  \leqno (5.8)$$
The Weyl symbol of ${\cal T}_h A $ is equal to the  Wick symbol of
$A$.

 }

\bigskip

{\it Proof.} Point a) is clear since $W_{Xh}$ is unitary. For any
$\theta$ in $[0, 1]$ define:
$$ T (\theta , h)  A = ( \pi h)^{-n} \int _{\R^{2n}} e^{ - {|X|^2 \over h} } W
_{\theta X , h } A W_{\theta X , h}^{\star } \ dX.  $$
Thus $T (1 , h) A = T_h A$ and $T(0, h)A = A$.  We verify that:
$$ {\partial \over \partial \theta }W
_{\theta X , h } A W_{\theta X , h}^{\star } = {i\over h } \sum
_{j=1}^n \Big [ x_j W_{\theta X , h} [P_j(h) , A] W_{\theta X ,
h}^{\star } - \xi_j W_{\theta X , h} [Q_j(h) , A] W_{\theta X , h}
^{\star} \Big ]. $$
Consequently:
$$ \Vert A -  {\cal T}_h A \Vert_{{\cal L}({\cal H}) } \leq
{1 \over h} \sum _{j=1}^n  ( \pi h)^{-n} \int _{\R^{2n}\times [0,
1]} e^{ - {|X|^2 \over h} } \Big [ |x_j|\  \Vert [P_j(h) , A]
\Vert_{{\cal L}({\cal H}) } + |\xi _j| \ \Vert [Q_j(h) , A]
\Vert_{{\cal L}({\cal H}) } \Big ] dx d\xi d\theta $$
$$ \leq  { C \over \sqrt h}  \sum _{j= 1}^n \Vert (ad P _j (h))   A
\Vert_{{\cal L}({\cal H}) } + \Vert (ad Q _j (h))   A \Vert_{{\cal
L}({\cal H}) } $$
proving point b) and also point c) with straightforward
modifications. For the point d) we see that:
$$ \sigma_h^{wick} ({\cal T}_h A )(X) = < ({\cal T}_h A ) \Psi _{Xh
} , \Psi _{X h} > = < ({\cal T}_h A ) W _{X , h} \Psi _{0h } ,  W
_{X , h} \Psi _{0 h} > $$
$$ = ( \pi h)^{-n} \int _{\R^{2n}} e^{ - {|Y|^2 \over h} }
< W _{Y , h } A W_{Y , h}^{\star }W _{X , h} \Psi _{0h } ,  W _{X ,
h} \Psi _{0 h} >  \ dY$$
$$ = ( \pi h)^{-n} \int _{\R^{2n}} e^{ - {|Y|^2 \over h} }
< A W _{X-Y , h} \Psi _{0h } ,W _{X-Y , h} \Psi _{0h } > dY . $$
We have used here (5.3). Consequently,
$$ \sigma_h^{wick} ({\cal T}_h A )(X)   = ( \pi h)^{-n} \int _{\R^{2n}} e^{ - {|Y|^2 \over h} }
\sigma_h^{wick } (A) (X-Y ) dY$$
which is  (5.8). According to proposition 4.4 we also have:
$$\sigma_h ^{wick}({\cal T}_h A) = e^{{h\over 4}\Delta} \sigma_h ^{weyl}({\cal T}_h A).  $$
Since the operator $e^{{h\over 4}\Delta}$ is one to one we then
deduce as it is mentioned,
$$   \sigma_h^{weyl} ({\cal T}_h A ) = \sigma_h^{wick}(A). $$

\hfill \carre

\bigskip

Next we consider the case of trace class operators without
additional assumptions. The result below does not have any
counterpart in the case of bounded operators.

\bigskip

{\bf Theorem 5.2.} {\it The space of operators written as
$OP_h^{weyl}(F)$ with $F$ in $W^{\infty 1}(\R^{2n})$ is dense in the
space  ${\cal L}^1 ({\cal H})$ of trace class operators.

}

\bigskip

{\it Proof.} For this purpose, we modify the approximation process
and we set:
$$ {\cal T}'_{\lambda} A = ( \pi \lambda)^{-n} \int _{\R^{2n}} e^{ - {|X|^2 \over \lambda} }
W _{X , 1} A W_{X , 1}^{\star } \ dX $$
for all
 $\lambda >0$ and for all trace class operators $A$. Let us show
 that:
$$\lim _{\lambda \rightarrow 0} \Vert {\cal T}' _{\lambda}(A)
- A \Vert _{{\cal L}^1({\cal H})}= 0 \leqno (5.9) $$
for all trace class operators $A$. Since we clearly have
$$ A = ( \pi \lambda)^{-n} \int _{\R^{2n}} e^{ - {|X|^2 \over \lambda} }
 A \ dX $$
then   we see that:
$$\Vert {\cal T}' _{\lambda}(A)
- A \Vert _{{\cal L}^1({\cal H})} \leq ( \pi \lambda)^{-n} \int
_{|X|< \delta }e^{ - {|X|^2 \over \lambda} } \Vert W _{X , h } A
W_{X , h}^{\star } - A \Vert _{{\cal L}^1({\cal H})} + ... $$
$$ ... + ( \pi \lambda)^{-n} \int _{|X|> \delta}e^{ - {|X|^2 \over \lambda}
} \Big [\Vert W _{X , \lambda } A W_{X , h}^{\star } \Vert _{{\cal
L}^1({\cal H})} + \Vert  A \Vert _{{\cal L}^1({\cal H})} \Big ] dX$$
for all $\delta >0$ and for all $\lambda >0$. For all trace class
operators $A$ and for all $\varepsilon
>0$ there exists $\delta >0$  such that:
$$ |X |< \delta ,  \Longrightarrow
 \Vert W _{X , 1
} A W_{X , 1}^{\star } - A \Vert _{{\cal L}^1({\cal H})} <
\varepsilon .  $$
Indeed, this property is first verified when $A$ is of the form
 $f \rightarrow < f , \varphi> \psi$ with $\varphi$ and $\psi$
in ${\cal S}(\R^n)$, it is next derived by density for finite-rank
operators and then, by density again for trace class operators.
Besides, $\delta >0$ being fixed, we have:
 $$ \lim _{\lambda \rightarrow 0 }( \pi \lambda)^{-n} \int _{|X|>\delta  }
 e^{ - {|X|^2 \over \lambda}
} \Big [\Vert W _{X , 1 } A W_{X , 1}^{\star } \Vert _{{\cal
L}^1({\cal H})} + \Vert  A \Vert _{{\cal L}^1({\cal H})} \Big ] dX =
0 . $$
The limit in (5.9) is then easily obtained. From (5.3), for all $X$
in $\R^{2n}$ we have:
$$W_{X, 1}{\cal T}' _{\lambda}(A)W_{X , 1}^{\star }  =
( \pi \lambda)^{-n} \int _{\R^{2n}} e^{ - {|Y|^2 \over \lambda} } W
_{X+Y , 1} A W_{X+Y , 1}^{\star } \ dY $$
$$ = ( \pi \lambda)^{-n} \int _{\R^{2n}} e^{ - {|X-Z|^2 \over \lambda} } W
_{Z , 1} A W_{Z , 1}^{\star } \ dZ . $$
Consequently, for all $\lambda >0$, the mapping $X \rightarrow W_{X,
1}{\cal T}' _{\lambda}(A)W_{X , 1}^{\star } $ is $C^{\infty} $ from
$\R^{2n}$ into ${\cal L}^1({\cal H})$. We see, setting $X = (x ,
\xi)$ as the variable of $\R^{2n}$,
$${\partial \over \partial x_j} W_{X, 1}{\cal T}' _{\lambda}(A)W_{X , 1}^{\star
}= - W_{X, 1} \Big [P_j(1) , {\cal T}' _{\lambda}(A)  \Big ] W_{X ,
1}^{\star } $$
$${\partial \over \partial \xi_j} W_{X, 1}{\cal T}' _{\lambda}(A)W_{X , 1}^{\star
}= W_{X, 1} \Big [Q_j(1) , {\cal T}' _{\lambda}(A)  \Big ] W_{X ,
1}^{\star } $$
where $P_j(1)$ is the operator of differentiation with respect to
$u_j$ and  $Q_j(1)$ is the multiplication operator by  $u_j$.
Consequently, all order iterated commutators   ${\cal T}'
_{\lambda}(A)$ with the position and momentum operators $P_j(1)$ and
$Q_j(1)$ are trace class. From the result of characterization of  C.
Rondeaux [R] (the analogous one of Beals's characterization for
trace class operators and recalled in the first part of this work)
it follows that ${\cal T}' _{\lambda}(A)$ is written as
$Op_1^{weyl}(F_{\lambda})$ with $F_{\lambda}$ in $W^{\infty
1}(\R^{2n})$ for all $\lambda
>0$.

\bigskip

{\bf 6. Proof of theorem  1.1.}

\bigskip

Let  $A$ be an operator in ${\cal L}({\cal H})$ satisfying the
assumptions in theorem 1.1. Set $F_h (X ) = \sigma _h^{wick} (A)
(X)$. This function is $C^{\infty}$ from proposition 4.1. Denoting
by $\varphi_t$ the Hamiltonian flow associated to the function $H(x
, \xi) = |\xi|^2 + V(x)$, we shall use the following function and
operators:
$$w_h (X , t) = F_h ( \varphi_t(X)) \hskip 2cm C_h(t ) = Op_h ^{weyl} (w_h (.,
t)).  \leqno (6.1) $$
We shall also use the operator ${\cal T}_h A$  appearing in section
5 and the following operator:
$$ B_h(t ) = e^{i{t\over h} \widehat H_h} {\cal T} _h(A) e^{- i{t\over h} \widehat
H_h}. \leqno (6.2) $$
We shall compare Wick symbols of the operators $A_h(t)$, $B_h(t)$
and $C_h(t)$ and then compare the Wick symbol of $C(t , h)$ with the
function $w_h(., t)$. This is accomplished in the three steps below.

\smallskip

{\it First step.} We have, from theorem 5.1,
$$ \Vert \sigma_h^{wick} ( A_h(t) -  B_h(t)) \Vert _{L^{\infty
}(\R^{2n})} \leq \Vert A_h(t ) - B_h(t) \Vert _{{\cal L}({\cal H})}
= \Vert A - T_h(A)\Vert _{{\cal L}({\cal H})} \leqno (6.3)$$
$$ ...  \leq C   \sqrt {h} I_h^{\infty} (A)$$
where $I_h^{\infty} (A)$ is defined in (1.13).

\smallskip

{\it Second step.} The comparison of  $B_h(t)$ and $C_h(t)$ comes
from Egorov's theorem. Nevertheless this requires some precisions
due to unusual estimates satisfied by the derivatives of $w_h(., t)$
that we first need to specify. Since $w_h(., 0) =
\sigma_h^{wick}(A)$ we deduce that:
$$ {\partial w_h (., 0) \over \partial x_j} = {i\over h} \sigma_h^{wick } (
[P_j(h) , A] ) \hskip 2cm {\partial w_h(., 0) \over \partial \xi_j}
= - {i\over h} \sigma_h^{wick } ( [Q_j(h) , A] ).  $$
Applying proposition 4.1 to the above commutators  we see  when $k
\geq 1$ that:
$$  \Vert  \nabla^k  w_h(., 0)
\Vert _{L^{\infty}(\R^{2n})} \leq C_{\alpha \beta } h^{1-(k+1 )/2}
 I_h^{\infty} (A).  $$
The derivatives of order $\geq 1$ of the Hamiltonian flow
$\varphi_t$ associated to the symbol $H(x , \xi) = |\xi|^2 + V(x)$
are bounded in $R^{2n }$ with a bound equaling to ${\cal O}(1+t^2)$.
Then there exists $M_k(t)$ such that:
$$  \Vert  \nabla ^k  w_h (., t)
\Vert _{L^{\infty}(\R^{2n})} \leq M_{k }(t) h^{1-(k+1 )/2}
I_h^{\infty} (A).  \leqno (6.4) $$
The operators $B_h(t)$ and $C_h(t)$ satisfy:
$$ -ih {\partial B_h(t) \over \partial t} = [ \widehat H_h , B(t , h)] \leqno (6.5)$$
$$ -ih {\partial C_h(t) \over \partial t} = -ih Op_h ^{weyl} (\partial _t w_h (., t))
= -ih Op_h ^{weyl} ( \{  H , w_h(., t) \}).   $$
From a standard result on the Weyl calculus recalled in proposition
3.2 of [AKN] (first part of this work), for all functions $F$ and
$G$ in $W^{\infty \infty}(\R^{2n})$, the operator $\widehat
R^{(2)}_h( F , G )$ defined by:
$$[ Op _h^{weyl} (F) , Op _h^{weyl} (F) ] = {h\over i} Op_h^{weyl} (
\{ F , G \} ) + \widehat R^{(2)}_h(  F , G )$$
satisfies the following estimate:
$$ \Vert \widehat R ^{(2)}_h (F , G) \Vert _{{\cal L}({\cal H})}
\leq C \sum _{ j\geq 2, k\geq 2 \atop  4 \leq j + k \leq   6n+8}
h^{(j+k)/2} \Vert \nabla ^j F \Vert _{L^{\infty}(\R^{2n})} \
 \Vert \nabla ^k G \Vert _{L^{\infty}(\R^{2n})}.  \leqno (6.6) $$
With these  notations, one may write:
$$-ih {\partial C_h(t)\over \partial t}  -  [ \widehat H_h , C_h(t )] =
  \widehat R^{(2)}_h(  H , w_h (., t) ).  \leqno (6.7)$$
Note that $\widehat R^{(2)}_h ( \Delta , w_h (., t) )=0$ and
consequently
$$ \widehat R^{(2)}_h ( H , w_h (., t) )   =  \widehat  R^{(2)}_h ( V , w_h (., t)
). $$
We then may apply inequality (6.6)  with the functions $F = V$ and
$G =w_h (., t)$. The inequality (6.6), those in (6.4) which are
verified by $w_h (., t)$, and the fact that all derivatives of $V$
are bounded allows us to write:
$$ \Vert \widehat R^{(2)}_h ( H , G_h (., t)) \Vert _{{\cal L}({\cal
H})} \leq M  (t)\ h^{3/2} \  I_h^{\infty} (A).  \leqno (6.8)$$
From theorem 5.1, the operator ${\cal T}_h A$ appearing in section 5
has a Weyl symbol equal to the Wick symbol of  $A$. Consequently,
the  Weyl symbol of $B_h(0) = {\cal T}_h (A)$ and the one of
$C_h(0)$ which is  $F_h = \sigma_h^{wick}(A)$ are equal. Thus,
$$B_h(0)= C_h(0). \leqno (6.9) $$
From (6.5), (6.7) and (6.9), Duhamel's principle implies:
$$ ih [ B_h(t ) - C_h(t ) ] = \int_0^t e^{ i {t-s \over h}
\widehat H_h}\ \widehat R^{(2)} _h (H , G_h (., t)) \ e^{ - i {t-s
\over h} \widehat H_h} \ ds . $$
Consequently, when $t>0$:
$$ \Vert  B_h(t) - C_h(t ) \Vert _{{\cal L}({\cal
H})}  \leq {1 \over h} \int_0^t \Vert \widehat R^{(2)}_h ( H , G_h
(., s)) \Vert _{{\cal L}({\cal H})}  ds.  $$
We then deduce that:
$$ \Vert \sigma_h ^{wick} \Big ( B_h(t) - C_h(t )  \Big ) \Vert
_{L^{\infty } (\R^{2n})} \leq \Vert  B_h(t ) - C_h(t) \Vert _{{\cal
L}({\cal H})} \leq M(t)   \sqrt {h}\  I_h^{\infty} (A).   \leqno
(6.10)
$$
{\it Third step.} From  proposition 4.4,
 $$ \sigma_h^{wick} ( (C_h(t)) = e^{{h\over 4}\Delta }
 \sigma_h^{weyl} ( C_h(t ) ) =  e^{{h\over 4}\Delta }
  w_h (., t). $$
Then
$$ \Vert \sigma_h^{wick} ( (C_h(t )) -
 w_h (., t)\Vert _{L^{\infty }(\R^{2n})} \leq {h\over 4}\int _0^1  \Vert
\Delta e^{{\theta h \over 4}\Delta } w_h(., t)\Vert _{L^{\infty
}(\R^{2n})} \leq \Vert  \Delta  w_h(., t)\Vert _{L^{\infty
}(\R^{2n})}.
$$
In view of the estimates  (6.4) satisfied by $ w_h (., t)$, we
obtain:
 $$\Vert  \sigma_h^{wick}  (C_h(t ))  -  w_h (., t) \Vert _{L^{\infty }(\R^{2n})}
 \leq M(t)   \sqrt {h}  I_h^{\infty} (A).  \leqno (6.11)$$
Since $w_h (. , t) = \Big ( \sigma_h^{wick } (A) \Big ) \circ
\varphi_t$,  then   inequality (1.14) in theorem 1.1 arises from
(6.3), (6.10) and (6.11).

\bigskip

{\bf 7. Proof of theorem  2.1.}

\bigskip

Let $\rho_h(t)$ be a family of solutions to the equation (TDHF)
(2.1),  satisfying the hypotheses of theorem 2.1.  Let $u_h(., t)$
be the function defined in (2.3). Let  $v_h (., t)$ be the solution
to Vlasov equation (2.5) such that $v_h(., 0) = u_h(., 0)$. We shall
use the functions $V_q (., \rho_h(t))$ and $V_{cl} (., v_h (., t))$
defined in (2.2) and (2.7) respectively. We shall also use the
following functions
$$ H_h ^{HF}(x , \xi, t)= |\xi|^2 +  V_q (x,
\rho_h(t))  \hskip 2cm H_h ^{VL}(x , \xi, t)= |\xi|^2 + V_{cl} (x,
v_h (., t))  \leqno (7.1) $$
and the associated operators through the Weyl calculus, namely:
$$ \widehat H_h ^{HF} = - h^2\Delta +  V_q (.,
\rho_h(t))  \hskip 2cm
 \widehat H_h ^{VL} = - h^2\Delta + V_{cl} (x,
v_h (., t)),   \leqno (7.2)$$
by using the same notation for the function and the corresponding
multiplication operator. We shall denote by  $w_h (X , t)$ the
solution to
$$ {\partial w_h \over \partial t} (. , t) = \{ H_h ^{HF}(. , t) ,
w_h(., t) \}, \leqno (7.3)$$
such that
$$ w_h(., 0) = v_h(., 0) = u_h(., 0).  \leqno (7.4)$$
In order to compare $v_h(., t)$ with $w_h(., t)$ we note that the
Vlasov equation (2.5) is written as
$$ {\partial v_h \over \partial t} (. , t) = \{ H_h ^{VL}(. , t) ,
v_h(., t) \}.  \leqno (7.5)$$
We shall use the operator $B_h(t)$ solution to
$$ih {d B_h(t) \over dt} =
  [\widehat H_h^{HF} (t) , B_h(t) ] \hskip 2cm B_h (0) = {\cal T}_h (\rho_h
  (0)), \leqno (7.6) $$
where ${\cal T}_h $ is the mapping used in section 5. Finally, we
shall also use the following operators:
$$ C_h(t) = (2 \pi h)^n  Op_h^{weyl} (w_h(., t)) \hskip 2cm
D_h(t) = (2 \pi h)^n  Op_h^{weyl} (v_h(., t)).  \leqno (7.7)$$
According to the point d) in theorem 5.1, we have
$$ \sigma_h^{weyl } (B_h(0) ) =\sigma_h^{weyl } \Big ( {\cal T}_h ( \rho_h(0)
) \Big ) = \sigma_h^{wick} ( \rho_h(0))  = (2\pi h)^n u_h(., 0).
\leqno (7.8)$$
Consequently,
$$ B_h(0) = C_h(0) = D_h(0) \leqno (7.9)$$
Theorem 2.1 is a consequence of the comparison between the Wick
symbol of the operators
 $\rho_h(t)$ and $B_h(t)$, between those of
$B_h(t)$ and $C_h(t)$,  between those of $C_h(t)$ and $D_h(t)$, and
finally between the Wick symbol of  $D_h(t)$ and the function
$v_h(., t)$. Each of these comparisons shall be written using the
expression $ I_h^{tr}( \rho_h(0))$ defined in (2.8), and shall
correspond to one step of the proof, but before that, we need three
more lemmata.

\bigskip

{\bf Lemma 7.1. }  {\it Let $\rho_h(t)$ be a family  of solutions to
the  (TDHF) equation  satisfying the assumptions in theorem 2.1. Let
$v_h (., t)$ be the function defined above. Then, for all integer
numbers  $k\geq 0$ we have
$$\Vert \nabla ^k v_h(., t) \Vert _{L^1 (\R^{2n})} \leq C_{k }(t) h^{-k/2}
\Vert \rho_h (0) \Vert _{{\cal L}^1({\cal H})}\leqno (7.10)$$
and for all integer numbers $k\geq 1$
$$ \Vert \nabla ^k v_h(., t) \Vert _{L^1 (\R^{2n})} \leq C_{k }(t) h^{-(k-1)/2}
I_h^{tr}( \rho_h(0)). \leqno (7.11)$$
These estimates remain valid when replacing the function $v_h(., t)$
by the function $w_h(., t)$.

 }

\bigskip

{\it Proof of the lemma.} Since we have (7.4) for $t=0$, then the
estimates (7.10) and (7.11) when $t=0$ come from proposition 4.2
applied with the operator $\rho_h(0)$ (for (7.10)) and with the
commutators $[P_j(h) , \rho_h(0)  ]$ and $[Q_j(h) , \rho_h(0)  ]$
(for (7.11)). Since the potentials  $V$  and $W$ are in $W^{\infty
\infty }(\R^n)$, then the functions $V_q (., \rho_h(t))$ and $V_{cl}
(., v_h (., t))$  are uniformly bounded together with all of their
derivatives. Consequently, the estimates satisfied at  $t=0$ by
$v_h(., 0) = w_h(., 0)$ remain valid along the Hamiltonian flows
associated to the two symbols $H_h^{HF}(., t)$ and $H_h^{VL}(., t)$.
Thus, the estimates (7.10) and (7.11) remain valid for all $t$ for
the function $v_h(., t)$ and the function $w_h(., t)$.

\bigskip

{\bf Lemma 7.2} {\it If $W$ is a function in $W^{\infty \infty }
(\R^{n})$, if we denote by   $W_x$ the multiplication by
$y\rightarrow W(x-y)$ and if  $A$ is trace class, then
$$e^{{h\over 4}\Delta_x }Tr ( W_x \circ A)= (2 \pi h)^{-n} \int _{\R^{2n} } W (x-y)
 \sigma _h ^{wick} (A) (y , \eta ) dy d\eta . \leqno (7.12) $$
}

\bigskip

{\it Proof of the lemma.} From (3.2),  if $W$ is as in the lemma and
if $A = Op_h^{weyl}(F)$ with $F$ in $W^{\infty 1}(\R^{2n})$, then we
have:
$$Tr ( W_x \circ A )= (2 \pi h)^{-n} \int _{\R^{2n} } W (x- y)
 \sigma _h ^{weyl} (A) (y , \eta ) dy d\eta $$
and then
$$e^{{h\over 4}\Delta_x }Tr ( W_x \circ A)
= (2 \pi h)^{-n} \int _{\R^{2n} } W (x-y) e^{{h\over 4}\Delta_x }
\sigma _h ^{weyl} (A) (y , \eta ) dy d\eta . $$
Besides
$$\int _{\R^{2n} } W (x-y)\ (  e^{{h\over 4}\Delta_{\eta} } -I)\ \sigma _h
^{weyl} (A) (y , \eta ) dy d\eta = 0, $$
and taking into account proposition 4.4 we deduce (7.12). Suppose
now that $A$ is an arbitrarily given trace class operator. Then
theorem 5.2 shows that there  exists a sequence of operators $A_j$,
written as $A_j = Op_h ^{weyl} (F_j)$ with $F_j$ in $W^{\infty
1}(\R^{2n})$ converging to $A$ in ${\cal L}^1({\cal H})$, ($h>0$
being fixed). Equality (7.12) valid for all the $A_j$ is also true
for $A $ when taking the limit while using proposition 4.2.
\hfill\carre

\bigskip

{\bf Lemma 7.3.} {\it With the above notations, we have:
$$ \Vert v_h(., t) - w_h (., t)  \Vert _{L^1 (\R^{2n})} \leq
C(t)  I_h ^{tr} ( \rho_h(0)) \left [ h + \int _{[0, t] } \Vert
v_h(., s) - u_h (., s) \Vert _{L^1 (\R^{2n})} ds  \right ].  \leqno
(7.13) $$
}

\bigskip

{\it Proof of the  lemma.} We deduce from (7.3)  and (7.5) that:
$${\partial (v_h - w_h ) \over \partial t} (. , t)  = \{ H_h^{HF}
(., t) , ( v_h(., t) - w_h(., t) ) \} + \{ (H_h ^{HF}(., t) - H_h
^{VL}(., t) ) , v_h (., t) \} $$
From  Duhamel's principle, since $v_h(. , 0) - w_h(., 0) = 0$ and
since the Hamiltonian flow associated to the function $H_h^{HF}(.,
t)$ preserves the norm of $L^1 (\R^{2n})$, we obtain
$$ \Vert v_h(., t) - w_h (., t)  \Vert _{L^1 (\R^{2n})} \leq
\int _{[0, t]} \Vert \{ H_h ^{HF} (., s) - H_h ^{VL} (., s) \ ,\ v_h
(., s)\} \Vert _{L^1 (\R^{2n})} ds . $$
We have:
$$ H_h ^{HF} (., s) - H_h ^{VL} (., s) =  ( I - e^{{h\over
4}\Delta_x })Tr ( W_x \circ \rho_h (t) ) + e^{{h\over 4}\Delta_x }Tr
( W_x \circ \rho_h (t)) - \int _{\R^{2n} } W (x-y) v_h (y , \eta ,
t) dy d\eta . $$
From lemma 7.2, we have:
$$ e^{{h\over 4}\Delta_x }Tr ( W_x \circ \rho_h (t)) =
\int _{\R^{2n} } W (x-y)\  u_h ( y , \eta , t) dy d\eta . $$
We then deduce
$$H_h ^{HF} (x , \xi, s) - H_h ^{VL} (x , \xi, s) =  ( I - e^{{h\over
4}\Delta_x })Tr ( W_x \circ \rho_h (t) ) + ... $$
$$ ... + \int _{\R^{2n} } W (x-y)\  \Big ( u_h ( y , \eta , t) -
v_h ( y , \eta , t) \Big ) dy d\eta . $$
In view of the preceding points,
$$ \Vert v_h(., t) - w_h (., t)  \Vert _{L^1 (\R^{2n})} \leq
C \int _{[0, t]}  \Vert \nabla v_h (., s)\} \Vert _{L^1 (\R^{2n})}
\Big [ h + \Vert u_h (., s) - v_h(., s) \Vert _{L^1 (\R^{2n})} \Big
] ds . $$
From lemma 7.1 (with $k=1$) we then deduce (7.13).

\bigskip

{\it End of the proof of  theorem 2.1. First step.} With the above
notations,  (TDHF) equation is written as:
$$ih {d \rho_h(t) \over dt} =
  [\widehat H_h ^{HF} (t) , \rho_h(t) ].   $$
We consequently have
$$ih {d (\rho_h(t) - B_h(t)) \over dt} =
  [\widehat H_h ^{HF} (t) , (\rho_h (t) - B_h(t)) ].   $$
Since the propagator associated to this equation preserves the trace
norm,  and since $B_h (0) = {\cal T}_h (\rho_h (0))$,  we then
deduce
$$ \Vert \rho_h(t) - B_h(t) \Vert _{{\cal L}^1({\cal H})} \leq \Vert
\rho_h  (0) -{\cal T}_h (\rho_h (0))\Vert _{{\cal L}^1({\cal H})} .
$$
Consequently, from the proposition 4.2 (with $m=0$) and theorem 5.1
(point c)),
$$ (2 \pi h)^{-n} \Vert \sigma_h ^{wick}
\Big (\rho_h (t) - B_h (t) \Big )\Vert _{L^1(\R^{2n})} \leq  \Vert
\rho_h  (0) -{\cal T}_h (\rho_h (0))\Vert _{{\cal L}^1({\cal H})}
\leq
 C  \sqrt h I_h^{tr}  (\rho_h (0)).   \leqno (7.14)$$
{\it Second step.} We shall bound  in norm $B_h(t)-C_h(t)$ and in
this aim we shall show that $B_h(t)$ and $C_h(t)$ verify similar
equations. The operator $B_h(t)$ verifies (7.6) whereas
$$ ih {d C_h(t) \over dt} = i h (2 \pi h)^n OP_h^{weyl} ( \{ H_h^{HF} (., t) ,
w_h(., t) \} ). $$
With the notations of  section 3 in [AKN] we have:
$$[ OP_h^{weyl} ( H_h^{HF}(., t) ) , OP_h^{weyl} ( w_h(., t) )] =
{h\over i} OP_h^{weyl}(\{  H_h^{HF} (., t) , w_h(., t) \} ) +
\widehat R_h^{(2)} ( H_h^{HF} (., t) , w_h(., t) ). $$
Consequently,
$$ih {d C_h(t) \over dt} - [\widehat H_h^{HF} (t) , C_h(t) ] =
(2 \pi h)^n \widehat R_h^{(2)} ( H_h^{HF} (., t) , w_h(., t) ).
\leqno (7.15) $$
We know that  $R_h^{(2)} ( F , G ) = 0$ for all function $G$ when
$F(x , \xi ) = |\xi|^2$. We can then replace $H_h^{HF}(., t)$ by
$V_q(., \rho_h(t))$ in the right hand-side of (7.15). By combining
(7.15) with equation (7.6) verified by $B_h(t)$ and
 using  Duhamel's principle and (7.4), we obtain
$$ \Vert B_h(t) - C_h(t) \Vert _{{\cal L}^1({\cal H})}  \leq {1
\over h} (2 \pi h)^n \int _{[0, t]} \Vert \widehat R_h^{(2)} ( V_q
(., \rho_h(s) ) , w_h(., s) )\Vert _{{\cal L}^1({\cal H})} ds . $$
From theorem 3.1 of [AKN] applied with $N= 2$, $F= V_q(., \rho_h(s))
$, $G= w_h(., s)$, $p= \infty$, $q=1$, we have
$$ \Vert  \widehat R_h^{(2)} (
V_q(., \rho_h(s)) , w_h(., s) )\Vert _{{\cal L}^1({\cal H})} \leq C
h^{-n} \sum_{ \alpha + \beta \leq 6n+8 \atop \alpha \geq 2 , \beta
\geq 2} h^{(\alpha + \beta )/2} \Vert \nabla ^{\alpha}V_q(.,
\rho_h(s)) \Vert _{L^{\infty }(\R^n)}\  \Vert \nabla ^{\beta }w_h(.,
s) \Vert _{L^{1 }(\R^{2n})}. $$
Since the potentials $V$ and $W$ are in $W^{\infty \infty }(\R^n)$
and since the $L^1(\R^{2n})$ norm of $w_h(., s)$  is bounded (lemma
7.1) then the derivatives of all order of $V_q(., \rho_h(s))$ are
bounded. For $\beta \geq 1$ the function $\nabla ^{\beta }w_h(., s)$
verifies the estimates (7.11) of lemma 7.1. Consequently, when $h\in
(0, 1]$,
$$ \Vert  \widehat R_h^{(2)} (
V_q(., \rho_h(s)) , w_h(., s) )\Vert _{{\cal L}^1({\cal H})} \leq
C(s) h^{3/2} I_h^{tr} ( \rho_h(0)). $$
Consequently,
$$\Vert B_h(t)  - C_h (t) \Vert _{{\cal L}^1({\cal H})}
\leq  C(t)   \sqrt h  I_h^{tr} (\rho_h(0)).
$$
We  then deduce (from the Proposition 4.2 with $m=0$) that:
$$ (2\pi h)^{-n} \Big \Vert \sigma_h^{wick} \big ( B_h(t) - C_h(t) \big
)\Big \Vert _{L^1(\R^{2n})} \leq C(t)   \sqrt h  I_h^{tr}
(\rho_h(0)).
 \leqno (7.16)$$
 {\it Third step.} From the proposition 4.4, we have
$$ (2 \pi h)^{-n} \Vert \sigma_h ^{wick} ( C_h(t) - D_h (t)) \Vert
_{L^1(\R^{2n})} \leq \Vert e^{{h\over 4}\Delta} (v_h (., t) - w_h(.,
t)  \Vert _{L^1(\R^{2n})}  \leq \Vert  (v_h (., t) - w_h(., t) \Vert
_{L^1(\R^{2n})} . $$
Then, from  lemma 7.2,
$$(2 \pi h)^{-n} \Vert \sigma_h ^{wick} ( C_h(t) - D_h (t)) \Vert
_{L^1(\R^{2n})} \leq C I_h ( \rho_h(0) )\left [ h +  \int_{[0, t]}
\Vert u_h(., s) - v_h(., s)\Vert _{L^1(\R^{2n})} ds \right ].
\leqno (7.17)$$
{\it Fourth step.} From  proposition 4.4 we have
$$  (2 \pi h)^{-n}\sigma_h ^{wick} ( D_h(t)  ) =
(2 \pi h)^{-n}e^{{h\over 4}\Delta } \sigma_h ^{weyl}  ( D_h(t)) =
e^{{h\over 4}\Delta }  v_h(., t). $$
Consequently, from the lemma 7.1 with $k=2$, we obtain
$$\Vert v_h(., t) - (2 \pi h)^{-n}\sigma_h ^{wick}
 ( D_h(t) \Vert _{L^1(\R^{2n})} \leq \Vert (e^{{h\over 4}\Delta
 } - I ) v_h(., t)\Vert _{L^1(\R^{2n})} \leq ...$$
 $$ ... \leq C h
  \Vert \nabla ^2  v_h(., t)\Vert _{L^1(\R^{2n})}
\leq  C(t)  \sqrt {h} I_h^{tr} ( \rho_h(0)).  \leqno (7.18)$$
Since $u_h(., t)$ is defined by (2.3), we obtain from estimates
(7.14), (7.16), (7.17) and (7.18) obtained in the four steps that:
$$\Vert  u_h(., t) - v_h(., t) \Vert _{L^1(\R^{2n})}
\leq C(t)  I_h^{tr}( \rho_h(0)) \left [ \sqrt {h} + \int_{[0, t]}
\Vert u_h(., s) - v_h(., s) \Vert _{L^1(\R^{2n})} ds \right ]. $$
From Gronwall's lemma, we have
$$\Vert  u_h(., t) - v_h(., t) \Vert _{L^1(\R^{2n})}
\leq C(t)\sqrt {h}   I_h^{tr} ( \rho_h(0))e^{C(t) I_h (\rho_h(0))}$$
 if $h\in (0, 1]$, with a different constant $C(t)$. Theorem 2.1 is complete.

\hfill\carre

\bigskip

 {\bf  8.  Approximative composition  of symbols with the
 Wick calculus.}

\bigskip

Being given two continuous operators $A$  and $B$  in ${\cal
S}(\R^n)$ we can define the  Wick symbols $\sigma_h^{wick}(A)$,
$\sigma_h^{wick}(B)$ and $\sigma_h^{wick}(A \circ B)$. When one of
these two symbols $\sigma_h^{wick}(A)$ or $\sigma_h^{wick}(B)$ is a
polynomial function, we can express $\sigma_h^{wick}(A \circ B)$
using an exact formula with the following notations. For any
function $F\in C^1(\R^{2n})$ we set
$$ \partial _j F ={1\over 2}\left [{\partial F\over \partial x_j} -
i{\partial F\over \partial \xi_j} \right ] \hskip 2cm \overline
{\partial} _j F ={1\over 2}\left [{\partial F\over \partial x_j} +
i{\partial F\over
\partial \xi_j} \right ] .  $$
For any multi-index $(\alpha , \beta)$ we set
$$ \partial ^{\alpha } \overline {\partial}^{\beta} = \partial
_1^{\alpha_1} ...\partial _n^{\alpha_n}  \overline {\partial
}_1^{\beta_1} ...  \overline {\partial }_n^{\beta_n} . $$
When $F$ is a function of several variables  in $\R^{2n}$ denoted by
$W$, $Y$, etc..., we shall write as a subscript the letter giving
the variable on which act the operator.

\bigskip

With these notations, if one of the  two symbols
$\sigma_h^{wick}(A)$ or $\sigma_h^{wick}(B)$ is a polynomial
function then  we have:
$$ \sigma _h^{wick} (A \circ B ) (X) = \sum _{|\alpha| < m}
 {(2h)^{\alpha}  \over  \alpha !}
\Big [\partial ^{\alpha} \sigma_h ^{wick}(A_h) (X) \Big ] \  \Big [
\overline {
\partial }^{\alpha} \sigma_h^{wick} (B_h) (X)\Big  ].  \leqno (8.1)$$
This formula justifies the terminology since it carries the Wick
order of creation and annihilation operators.

\bigskip

In this section, we shall give a similar  formula when one of the
two operators  is trace class while the other one  associated
through the Weyl calculus has a symbol in $W^{\infty \infty
}(\R^{2n})$. In this case, for all integer numbers $m\geq 1$ and for
all $h>0$ we shall denote by $ R_m (A , B; . ; h) $ the function on
$\R^{2n}$ defined by the equality
$$ \sigma _h^{wick} (A \circ B ) (X) = \sum _{|\alpha| < m}
 {(2h)^{\alpha}  \over  \alpha !}
\Big [\partial ^{\alpha} \sigma_h ^{wick}(A_h) (X) \Big ] \  \Big [
\overline {
\partial }^{\alpha} \sigma_h^{wick} (B_h) (X)\Big  ] + R_m (A , B;X ; h) . \leqno (8.2) $$

\bigskip

The main result of this section is the following one.

\bigskip

{\bf Theorem 8.1.} {\it We consider two operators $A$ and $B$ such
that

\smallskip

1. The operator $A$ is written as $A= Op_h^{weyl} (F)$ where $F\in
W^{\infty \infty  }(\R^{2n})$.

\smallskip

2. The operator $B$ is trace class.

\smallskip

Then, for all integer numbers $m \geq 1$ there exists a constant
$C_m>0$ (depending only on $m$ and $n$) such that the functions $R_m
(A , B;X ; h)$ defined by the equality (8.2)  and the analogous
function  $R_m (B , A;X ; h)$ verify:
$$(2 \pi h)^{-n}  \int _{\R^{2n} } \Big | R_m (A , B;X ; h)\Big | \ dX
\leq  C_{mn}  \     \ \Vert B_h\Vert _{{\cal L}^1({\cal H})}  \ \sum
_{\alpha \in E_{mn} }  h^{|\alpha|/2}\Vert
\partial ^{\alpha} F \Vert _{L^{\infty} (\R^{2n})}, \leqno (8.3)$$
where $E_{mn}$ is the following set of multi-indexes,
$$ E_{mn} = \{ \alpha \in \N^n , \ \ \ \ m \leq |\alpha| \leq \sup (
m, n+1) \}, \leqno (8.4) $$
and similarly for $R_m (B , A;X ; h)$.

}

\bigskip

We shall use the function $S_h(A)$ defined in (4.2) and those which
are similarly associated to $B$ and $A \circ B$. We shall refer as
$S_h(A)$ for the bi-Wick symbol of $A$. Using (4.5) we have:
$$ \sigma_h^{wick} (A \circ B) (X ) = (2 \pi h)^{-n} \int _{\R^{2n}}\
{\cal B}_h (X , Y  , Y , X) \ S_h (B) (X , Y) \ S_h (A) (Y , X) \
dY.  \leqno (8.5) $$
The proof of theorem 8.1 relies on a Taylor expansion for the
bi-Wick symbol of the operator $A_h= Op_h^{weyl} (F)$ in a
neighborhood of the diagonal. It is standard that this bi-symbol is
holomorphic in $X= x + i \xi$ and anti-holomorphic in $Y$. We can
give a quick proof. Iterating (4.5) we see that
$$ S_h(A) (X , Y) = (2 \pi h)^{-n} \int _{\R^{4n}} {\cal B}_h ( X ,
U , V , Y) S_h(A) (U , V) dU dV, \leqno (8.6)$$
where ${\cal B}_h$ is defined in (4.8). Replacing in (4.8) the
expression (4.3) giving the scalar product of two coherent states,
we obtain an explicit expression  of the function ${\cal B}_h ( X ,
U , V , Y) $, showing that it is holomorphic in $X$,
anti-holomorphic in $Y$ and $S_h(A)$ inherits therefore of these
properties. From these points,
$$ S_h (A) (X , Y) = \sum _{\alpha \in \N ^n }  { (X-Y)^{\alpha} \over
\alpha!}  (\partial ^{\alpha} \sigma_h ^{wick} (A) )(Y)= \sum
_{\beta \in \N^n } { (\overline Y- \overline X)^{\beta} \over
\beta!}( \overline {\partial} ^{\beta} \sigma_h^{wick} (A)) (X).
$$
The first step  in the proof of theorem 8.1 is an uniform estimation
as $h$ tends to $0$ of the remaining terms of order $m$ in the above
expansions. Namely,
$$ S_h (A) (X , Y) = \sum _{|\alpha |<m} { (X-Y)^{\alpha} \over
\alpha!} (\partial ^{\alpha} \sigma_h ^{wick} (A) )(Y)
 + R_{m}
(A; X , Y; h), \leqno (8.7)$$
$$ S_h (A) (X , Y) = \sum _{|\beta|<m} { (\overline Y- \overline
X)^{\beta} \over \beta!} ( \overline {\partial} ^{\beta}
\sigma_h^{wick} (A)) (X) + \widetilde R_{m} (A ;X , Y; h). \leqno
(8.8)$$
We shall restrict ourselves to the case where $A=  Op_h^{weyl}(F)$
is a pseudo-differential operator.

\bigskip

{\bf Proposition 8.2.} {\it Let $A$ be an operator of the form $A=
Op_h^{weyl}(F)$ where $F\in W^{\infty \infty } (\R^{2n})$. For all
integer number $m \geq 1$, let $R_{m} (A ; X , Y; h) $ and
$\widetilde R_{mh} (A ; X , Y; h) $ be the functions defined in
(8.7) and (8.8). Then, there exists a function $G_m$  in
$L^1(\R^{2n})$ such  that, for all $h$ in $(0, 1]$
$$e^{- {1\over 4h}|X-Y|^2} | R_{m}(A ; X , Y; h)|    \ \leq  \
\  \ \ G_m \left ({X - Y\over \sqrt {h}} \right ) \sum _{ \alpha \in
E_{mn}} h^{|\alpha|/2} \Vert \partial ^{\alpha} F\Vert
_{L^{\infty}(\R^{2n})}, \leqno (8.9)$$
where $E_{mn}$ is defined in (8.4). An analogous estimate remains
valid for the function $\widetilde R_{m} (A ; X , Y; h)$ defined in
(8.8).

 }

\bigskip

{\it Proof.} Proposition 4.4 gives an expression of the Wick symbol
of  $A_h = Op_h^{weyl}(F)$
$$ \sigma_h^{wick} (A_h) (X) = (\pi h)^{-n} \int
_{\R^{2n}}e^{-{|X-Z|^2 \over h}} F(Z) dZ $$
Since $S_h (A_h) (X , Y)$ is a function holomorphic in $X$,
anti-holomorphic in $Y$ and equaling to $\sigma_h^{wick}(A_h)$ on
the diagonal, we  necessarily have
$$  S_h (A_h) (X , Y) = (\pi h)^{-n} \int
_{\R^{2n}}e^{ -{1\over h}  (Z - X )  .(\overline Z -  \overline Y)}
F(Z) dZ . $$
In order to derive  the  Taylor expansion, we set for all $\theta
\in [0, 1]$
$$F_h( \theta  , X , Y) = (\pi h)^{-n} \int _{\R^{2n}} F(Z) e^{{1 \over
h} \varphi (\theta  , X , Y , Z) } dZ, $$
$$ \varphi (\theta  , X , Y , Z) =  - \Big (Z-Y - \theta (X - Y) \Big ) .(\overline
Z -  \overline Y) dZ. $$
Thus, we have $F_h( 1 , X , Y) = S_h(A_h) (X , Y)$ and $F_h( 0 , X ,
Y) = S_h(A_h) (Y , Y)$. The Taylor expansion is then written as
$$S_h(A_h) (X , Y) = \sum _{k < m} {1 \over k!} \partial _{\theta}^k F_h
(0, X , Y ) + R_{m} (A_h ; X , Y; h), $$
$$ R_{m} (A_h ; X , Y; h) =  \int _0^1 {(1-\theta )^{m-1}  \over (m-1)!}
\partial _{\theta} ^m F_h (\theta , X , Y) \ d\theta .   $$
Differentiating the above equality through the integral and
integrating by parts, we obtain:
$$ \partial _{\theta}^k F_h (\theta , X , Y ) =  (\pi h)^{-n} \int _{\R^{2n}}  e^{{1 \over
h} \varphi (\theta  , X , Y , Z) }\ \Big [ (X-Y). \partial _Z\Big
]^k F(Z) \ dZ . $$
Consequently,
$${1 \over k!}  \partial _{\theta}^k F_h (\theta , X , Y ) =  \sum _{|\alpha|=k}
{ (X-Y)^{\alpha}\over \alpha !} (\pi h)^{-n} \int _{\R^{2n}}  e^{{1
\over h} \varphi (\theta  , X , Y , Z) }\ \Big ( \partial ^{\alpha}
F\Big ) (Z) \ dZ . $$
In particular, for $\theta =0$,
$$  (\pi h)^{-n} \int _{\R^{2n}}  e^{{1
\over h} \varphi (0 , X , Y , Z) }\ \partial _Z^{\alpha} F(Z) \ dZ =
 (\pi h)^{-n} \int _{\R^{2n}}  e^{{1
\over h} |Y - Z|^2 }\ \Big ( \partial ^{\alpha} F\Big ) (Z) \ dZ $$
$$ ... = \Big ( e^{{h\over 4}\Delta }\partial ^{\alpha}  F\Big )  (Y) =
\Big ( \partial ^{\alpha} \sigma_h ^{wick} (A)\Big )  (Y)  . $$
The  Taylor  remaining term is given  by:
$$R_{m} (A_h ; X , Y; h) = m \sum _{|\alpha|=m}
{ (X-Y)^{\alpha}\over \alpha !}  (\pi h)^{-n}  \int _{[0, 1] \times
\R^{2n} } (1-\theta )^{m-1}  e^{{1 \over h} \varphi (\theta  , X , Y
, Z) }\
\partial ^{\alpha} F(Z) \ d\theta  dZ . $$
Consequently,
$$\big |R_{m} (A_h ; X , Y; h)\big | \ \leq \ m |X-Y|^m \left [ \sum _{|\alpha|=m}
\Vert \partial ^{\alpha} F \Vert _{L^{\infty}} \right ]  \ (\pi
h)^{-n} \int _{[0, 1] \times \R^{2n} } (1-\theta )^{m-1}  e^{{1
\over h} {\rm Re} \varphi (\theta  , X , Y , Z) }\ d\theta  dZ . $$
We see that:
$${\rm Re} \varphi (\theta  , X , Y , Z) = - \left | Z-Y - {\theta \over 2} ( X-Y)\right |^2 +
\ {\theta ^2\over 4} |X-Y|^2 . $$
If $m\geq n+1$ and $C>0$ the following function
$$ G_m (X ) = C \ |X |^m \int _0^1 (1-\theta )^{m-1} \ e^{ {\theta ^2 -1 \over
4} |X |^2} d\theta $$
is in $L^1(\R^{2n})$. We can find $C>0$ such that (8.9) is
satisfied.  When $m\leq n $ we have
$$R_{m}(A_h ; X , Y; h) = R_{n +1 }(A_h ; X , Y; h) + \sum _{ m\leq  |\alpha| \leq n }
{ (X-Y)^{\alpha} \over \alpha!} \Big (\partial_X ^{\alpha} \sigma_h
^{wick}(A) \Big ) (Y) . $$
When $m \leq n$  we can then  find $C>0$ such that the inequality
(8.9) is verified when setting
$$G_m (X) = G_{n+1}  (X ) + C \ e^{-{1\over 4}|X|^2} \ \sum _{k=m}^{n}
|X|^k $$
This function $G_m$ is also in $L^1(\R^{2n})$. In both of the two
cases we have the estimation (8.9). The estimation concerning
$\widetilde R_{m} (A_h ; X , Y; h)$ is similarly proved.

\hfill \carre

\bigskip

{\it End of the proof of theorem 8.1.} Using (8.5) and  using  the
asymptotic expansion in the first  variable (here $Y$) of the
function $S_h(A_h) (Y , X)$ given in the proposition 8.2 when $A_h =
Op_h^{weyl} (F)$, we  obtain:
$$ \sigma _h^{wick}(A \circ B) (X) = \sum_{|\alpha|< m} {a_{\alpha } (X , h) \over \alpha
!}  \partial^{\alpha} \sigma_h^{wick}(A) (X)+ R_m (A , B;X ; h)$$
where
$$ a_{\alpha} (X , h) = (2 \pi h)^{-n} \int _{\R^{2n}}\
{\cal B}_h (X , Y  , Y , X)  \ S_h (B) (X , Y) (Y-X)^{\alpha} dY$$
$$R_m (A , B;X ; h) = (2 \pi h)^{-n} \int _{\R^{2n}}\ {\cal B}_h (X , Y  , Y , X)
  \ S_h (B) (X , Y) R_{m} (A; X , Y; h) \ dY . $$
From  (4.9) we have:
$${\cal B}_h (X , Y  , Y , X) = {\cal B}_h (X , X  , Y , X) =
e^{ -{1 \over 2h}(X - Y) ( \overline X - \overline Y)} . $$
Therefore we obtain that
$$a_{\alpha} (X , h)
= (2h)^{|\alpha|} (2\pi h)^{-n} \int _{\R^{2n}}S_h (B) (X , Y)
\overline {\partial }_X ^{\alpha} {\cal B}_h (X , Y , Y , X )   \ dY
$$
$$ =(2h)^{|\alpha|} (2\pi h)^{-n} \overline {\partial
}_X ^{\alpha}\int _{\R^{2n}} {\cal B}_h (X , X , Y , X )S_h (B) (X ,
Y) \ dY$$
$$ = (2h)^{|\alpha|}  \overline {\partial }^{\alpha} \sigma_h^{wick}
(B) (X) . $$
We have used the fact that $\overline {\partial }_X S_h(B) (X , Y) =
0$ and the fact that the bi-symbol $S_h(B)$ verifies the property
(8.6). Besides, from the bound of $R_{m} (A; X , Y; h) $ given by
proposition 8.2,
$$ |R_m (A , B;X ; h)| \leq  C_m(  F, h)\
 \ (2 \pi h)^{-n} \int _{\R^{2n}}\ e^{ -{1 \over 4h}|X-Y|^2}
   \ |S_h (B) (X , Y)|
   G_m  \left (  {X-Y\over \sqrt {h}} \right )  dY , $$
$$ C_m (F, h)=   \sum _{ \alpha  \in E_{mn }}  h^{|\alpha|/2}
\Vert \partial ^{\alpha} F \Vert _{L^{\infty}} , $$
where $G_m$ is a function in $L^1(\R^{2n})$. Consequently,
$$(2 \pi h)^{-n}  \int _{\R^{2n} } \Big | R_m (A , B;X ; h)\Big | \ dX
\leq ... \hskip 7cm \ $$
$$\ \hskip 1cm  ... \leq C_m (F , h)\   (2 \pi h)^{-2n} \int _{\R^{4n}}  e^{ -{1 \over
4h}|X-Y|^2}
   \ |S_h (B) (X , Y)|
   G_m  \left (  {X-Y\over \sqrt {h}} \right )  dX dY  . $$
From  proposition 4.3, the point concerning $A\circ B$ is proved and
the one concerning $B\circ A$ is similarly derived.

\hfill \carre

\bigskip

{\bf 9. The equation satisfied by the Wick symbol of a solution to
(TDHF).}

\bigskip

If one only supposes  that $\rho_h(t)$ is a classical solution to
(TDHF)à without assuming any additional hypothesis,  then the  Weyl
symbol of $\rho_h(t)$ is a continuous function  on $\R^{2n}$ with
derivatives understood in the sense of
 distributions. However,
from  proposition 4.2 it is noted that its Wick symbol is in
$W^{\infty 1}(\R^{2n})$ and therefore it is $C^{\infty}$. We shall
prove in this section that the Wick symbol satisfies a differential
equation, with an asymptotic expansion in powers of $h$, the first
term of the equation being the Vlasov equation, and the error term
being as small as wanted, in terms of power of $h$, and of
$L^1(\R^{2n})$. For the introduction ofsuch an equation, see
[DLERS].

\bigskip

{\bf Proposition 9.1.} {\it a) Let $\rho_h (t)$ be a classical
solution to (TDHF). For all $t\in \R$ set:
$$ \varphi _h (X , t) = (2 \pi h)^{-n} \ \sigma_h ^{weyl } (\rho_h (t))\leqno (9.1) $$
$$u_h (X , t) = (2 \pi h)^{-n} \ \sigma_h ^{wick } (\rho_h (t)) \leqno (9.2) $$
Then,

 \smallskip

a) We have, in sense of  distributions on $\R^{2n} \times \R$,
$$ {\partial \varphi _h \over \partial t }  + 2 \sum _{j=1}^n \xi_j {\partial
\varphi _h \over \partial x_j}   = {1 \over ih}(2\pi h)^{-n}
\sigma_h^{weyl} \Big ([ V_q (\rho_h (t) ) ,\rho_h (t) ] \Big  ).
\leqno (9.3)$$
b) We have:
$$ {\partial u_h \over \partial t }  + 2 \sum _{j=1}^n \xi_j {\partial
u_h \over \partial x_j} + h \sum _{j=1}^n  {\partial^2 u_h \over
\partial x_j \partial \xi_j}  = {1 \over ih}(2\pi h)^{-n} \sigma_h^{wick} \Big ([ V_q
(\rho_h (t) ) ,\rho_h (t) ] \Big  ).  \leqno (9.4)  $$

}

\bigskip

We observe that when the potentials are turned off  ($V=W=0$) then
the Weyl symbol  of $\rho_h (t)$ is the only one that exactly
verifies  Vlasov equation (but in the sense of distributions).

\bigskip

{\it Proof. Point a) } Let $\rho_h(t)$ be a classical solution of
(TDHF). Let $u_h (., t)$ be the function defined in (9.2). Let
$\psi$ be a $C^{\infty }$ function  with compact support in $\R^{2n}
\times \R$. For all $t\in \R$, set $A_h(t) = Op_h^{weyl} ( \psi (.,
t))$. The mapping $A_h$ is  $C^1$ from $\R$ into ${\cal L}^1 ({\cal
H})$ and it is continuous from $\R$ into ${\cal D}$. We shall
represent the integral
$$ I_h = (2 \pi h)^{-n} \int _{\R^{2n} \times \R} \varphi_h (x , \xi, t)
\left [ {\partial \psi  \over \partial t }  + 2 \sum _{j=1}^n \xi_j
{\partial \psi  \over \partial x_j}(x , \xi, t) \right ] dx d\xi dt
.
$$
We know that:
$$ \sigma_h^{weyl}( h [ \Delta , A_h(t)] )(x , \xi)  = 2 i \sum _{j=1}^n \xi_j
{\partial \psi  \over \partial x_j}(x , \xi, t) . $$
Consequently,   we have from (3.2),
$$ I_h = \int _{\R} Tr \Big ( \rho_h (t) \circ \big ( A'_h(t) - i h [ \Delta
, A_h(t)] \big ) \Big )  dt .
$$
Since the operators $\rho_h (t) $ and $A_h(t)$ are in ${\cal D}$, we
have:
$$ Tr \big ( \rho_h (t) [\Delta , A_h (t) ] \big ) = - Tr \big ( [ \Delta , \rho_h (t)]  A_h (t)
\big ). $$
Consequently,
$$ - I_h = \int _{\R} Tr \Big ( \big ( \rho'_h (t) - i h [ \Delta ,
\rho_h (t) ] \big ) \circ A_h(t) \Big ) dt$$
$$ ={1 \over ih }   \int _{\R} Tr \Big ( [ V_q ( \rho_h (t)) ,
\rho_h (t) ] \circ A_h (t) \Big ) dt . $$
Using  (3.2) again,
$$- I_h ={1 \over ih } (2 \pi h)^{-n}  \int _{\R^{2n} \times \R}
\sigma_h^{weyl}\Big (  [ V_q ( \rho_h (t)) , \rho_h (t) ]\Big )
\varphi (x , \xi , t) dx d\xi dt . $$

\smallskip

 {\it Point b)} Let $\rho_h(t)$ be a classical solution to (TDHF).
 Let $u_h (., t)$ and $\varphi_h(., t)$ be the
functions defined in (9.1) and (9.2). We know (see proposition 4.4),
that $u_h (., t) = e^{{h\over 4} \Delta} \varphi _h(., t)$. Applying
the operator $e^{{h\over 4} \Delta}$ to the two hand-sides of (9.3)
we obtain (9.4). \hfill\carre

\bigskip

Combining theorems 9.1 together with 8.1 and lemma 7.2 together with
proposition 4.4 we directly obtain  an equation satisfied by the
Wick symbol of a solution to (TDHF) without any supplementary
hypothesis and  with  an arbitrary high order of accuracy. However,
without supplementary hypothesis, this equation does not allow to
obtain  an asymptotic expansion of the Wick symbol.

\bigskip

{\bf Theorem 9.2.} {\it Let $V$ and $W$  be two potentials in
$W^{\infty \infty }(\R^n)$. Let $\rho_h (t)$ be a classical solution
to (TDHF) such that $\rho_h (0) \geq 0$. Let $u_h (. , t)$ be the
function defined in (2.3). For all functions $f$ in $L^1(\R^{2n})$,
set:
$$ \Phi_h ( f ) (x) = ( e^{{h\over 4} \Delta} V) (x) +  \int _{\R^{2n}} W (x
- y ) f(y , \eta) dy d\eta . $$
Then the function $u_h$ verifies, for all integer numbers $m\geq 2$:
$$ {\partial u_h \over \partial t }  + 2 \sum _{j=1}^n \xi_j
{\partial u_h \over \partial x_j} + h \sum _{j=1}^n  {\partial^2 u_h
\over
\partial x_j \partial \xi_j}  = ... \hskip 8cm   $$
$$ ... =  {1 \over ih}\sum _{1\leq |\alpha| < m}
 {(2h)^{\alpha}  \over  \alpha !} \Big [\partial ^{\alpha} \Phi_h ( u_h (.,
 t)) \ \overline { \partial }^{\alpha} u_h (., t) \ -\ \partial ^{\alpha}
w_h (., t) \  \overline { \partial }^{\alpha} \Phi_h ( u_h (.,
 t))  \Big ] + R_m (. , t , h)$$
where the function $R_m(., t, h)$ is in
 $L^1(\R^{2n})$ for all $t\in \R$. For all $T>0$, there exists $C_m(T)>0$ such
 that:
$$ (2 \pi h)^{-n} \int _{\R^{2n} } \Big | R_m (X ; h)\Big | \ dX
\leq C_m (T)  h^{{m\over 2}-1 } .  $$
}

\bigskip

{\bf 10. The analogue of Ehrenfest's time.}

\bigskip

The  below result shows that the limit in corollary 2.2  cannot be
uniform on $\R$, even without potentials $V$ and $W$ entirely
vanishing, since exchanging there the two limits is false. In other
words, according to the terminology of [BR], [dBR], the analogue of
the Ehrenfest time for collary 2.2 is finite.

\bigskip

{\bf Theorem 10.1.} {\it  Let $h>0$ be fixed. Let $\rho_h(t)$ be a
classical solution to the (TDHF) equation  corresponding to $V=W=0$.
We suppose that $\rho_h(0) \geq 0$ and assume that the trace of
$\rho_h(0)$ equals 1. Let $u_h(., t)$ and $v_h(. , t)$ be the
functions defined in section 2. Then we have:
$$ \lim _{t\rightarrow \pm \infty } \Vert u_h (., t) - v_h (., t)
\Vert _{L^1(\R^{2n})} = 2.  \leqno (10.1)$$

}

\bigskip

We recall here that the two functions $u_h(., t)$ and $v_h(., t)$
are in the unit ball of $L^1(\R^{2n})$.

\bigskip

{\it Proof.} When $V=W=0$ the function $u_h(., 0)$ verifies from the
proposition 9.1,
$$ {\partial u_h \over \partial t }  + 2 \sum _{j=1}^n \xi_j {\partial
u_h \over \partial x_j} + h \sum _{j=1}^n  {\partial^2 u_h \over
\partial x_j \partial \xi_j}  = 0.   \leqno (10.2)  $$
The function $v_h (., t)$ is the solution to the Vlasov equation
equaling to $u_h(., 0)$ for $t=0$. Then,   $ v_h (x , \xi , t) = u_h
( x - 2t\xi , \xi, 0)$ in the case of vanishing potentials. Set
$$ U_h(x , t) = \int _{\R^n} u_h ( x + 2t \xi , \xi , t ) d\xi.  \leqno (10.3) $$
We have:
$$\Vert u_h (., t) - v_h (., t)
\Vert _{L^1(\R^{2n})} = \int _{\R^{2n}} | u_h (x + 2 t \xi , \xi, t)
- u_h (x + 2 t \xi , \xi, t) | dx d\xi \geq\int _{\R^n} | U_h ( x ,
t) - U_h (x , 0)| dx .  $$
From (10.2) and (10.3),
$$ {\partial U_h \over \partial t} (x , t) = -h \int _{\R^n} \sum _{j=1}^n
{ \partial^2 u_h \over \partial x_j \partial \xi_j} (x + 2t \xi ,
\xi, t) d\xi = 2ht \Delta_x U_h(x , t) .
$$
Consequently,
$$ U_h(. , t) = e^{ ht^2 \Delta_x} U_h (., 0) . $$
It is standard that:
$$ \lim _ {\lambda \rightarrow +\infty } \Vert e^{ \lambda
\Delta_x}F - F\Vert _{L^1(\R^{n})} = 2\Vert F \Vert _{L^1(\R^{n})},
$$
if  a function $F\geq 0$ is in $L^1 (\R^n)$. The proof of the
theorem is therefore completed.

\hfill\carre

\bigskip

\centerline{\bf  References}

\bigskip

[AN1] Z. Ammari, F. Nier, {\it Mean field limit for bosons and
infinite dimensional phase space analysis}, Ann. Henri Poincar\'e,
{\bf 9} (2008) 8, 1503-1574.

\medskip

[AN2] Z. Ammari, F. Nier, {\it Mean field limit for bosons and
propagation of Wigner measures}, J. Math. Phys, {\bf 50} (2009), 4,
042107.

\medskip

[AKN] L. Amour, M. Khodja et J. Nourrigat {\it The classical limit
of the time dependent
  Hartree-Fock equation. I. The Weyl symbol of the solution}.
  Preprint (2011).

\medskip

[B] R. Beals, {\it Characterization of pseudo-differential operators
and applications},  Duke Math. J.,{\bf 44} (1977),1, 45-57.

\medskip

[BR] A. Bouzouina, D. Robert, {\it Uniform semiclassical estimates
for the propagation of quantum observables,} Duke Math. J, {\bf 11},
2 (2002), 223-252.

\medskip

[BdPF1]  A. Bove, G. da Prato, G. Fano, {\it An existence proof for
the Hartree-Fock time-dependent problem with  bounded two-body
interaction}, Comm. Math. Phys, {\bf 37} (1974), 183-191.

\medskip

[BdPF2]  A. Bove, G. da Prato, G. Fano, {\it On the Hartree-Fock
time-dependent Problem,} Comm. Math. Phys, {\bf 49} (1976) 25-33.

\medskip

[CV] A.-P.Calder{\`o}n, R. Vaillancourt, {\it A class of bounded
pseudo-differential operators}, Proc. Nat. Acad. Sci. U.S.A., {\bf
69} (1972), 1185-1187.

\medskip

[CTDL] C. Cohen-Tannoudji, B. Diu, F. Lalo\"e, {\it M\'ecanique
quantique}, Hermann, 1973.

\medskip

[CR1] M. Combescure, D. Robert, {\it Semiclassical spreading of
quantum wave packets and applications near unstable fixed points of
the classical flow, } Asymptotic Anal, {\bf 14} (1997) 377-404.

\medskip

[CR2] M. Combescure, D. Robert, {\it Quadratic quantum Hamiltonians
revisited}, Cubo ({\bf 8}) (2006), 3, 61-86.

\medskip

[CF] A. Cordoba, C. Fefferman, {\it Wave packets and Fourier
Integral operators,} Comm. in P.D.E, 3 ({\bf 11}) (1978), 979-1005.

\medskip

[dBR] S. de Bi\`evre, D. Robert, {\it Semiclassical propagation and
the $log ( h^{-1})$ barrier. }  Int. Math. Res. Not. {\bf 12}
(2003), 667-696.

\medskip

[DLERS] A. Domps, P. L'Eplattenier, P.G. Reinhard, E. Suraud, {\it
The Vlasov equation for Coulomb systems and the Husimi picture,}
Annalen der Physik, {\bf 6} (1997), 455-467.

\medskip

[L1] N. Lerner, {\it Some facts about the Wick calculus,} in {\it
Pseudodifferential Operators,} 135-174, Lecture Notes in Math. 1949,
Springer, Berlin 2008.

\medskip

[R1] D. Robert, {\it Autour de l'approximation semiclassique},
Progress in Mathematics, {\bf 68}, Birkhäuser Boston, Inc., Boston,
MA, 1987.

\medskip
[R2] D. Robert, {\it  Propagation of coherent states in quantum
mechanics and applications,}  Partial differential equations and
applications, 181–252, Sémin. Congr, 15, SMF, Paris, 2007.

\medskip

[R] C. Rondeaux, {\it Classes de Schatten d'op\'erateurs
pseudo-diff\'erentiels,} Ann. E.N.S, {\bf 17}, 1, (1984), 67-81.

\medskip

\medskip
[S] R. Schubert, {\it Semiclassical localization in phase space},
PhD thesis, Ulm, 2001.

\medskip

[U1] A. Unterberger, {\it Oscillateur harmonique et op\'erateurs
pseudo-diff\'erentiels}, Ann. Inst. Fourier, XXXIX, 3, 1979.

\medskip

[U2] A. Unterberger, {\it  Les op\'erateurs m\'etadiff\'erentiels},
Lecure Notes in Physics {\bf 126}, 205-241 (1980).

\vskip 2cm

laurent.amour@univ-reims.fr

\bigskip

khodja@univ-reims.fr

\bigskip

jean.nourrigat@univ-reims.fr

\end